\documentclass[a4paper,11pt,aps,prd,superscriptaddress,nofootinbib,tighten,preprint]{revtex4}

\pdfoutput=1
\usepackage{graphicx,floatflt,amssymb,epsf,rotate} 
\usepackage{ulem}
\usepackage[utf8]{inputenc}
\usepackage[english]{babel}
\usepackage{amsmath}
\usepackage{subfigure}
\usepackage{amsfonts}
\usepackage{amssymb}
\usepackage{color}
\usepackage{cancel}
\usepackage{graphicx}
 
\usepackage{amsmath,amsfonts,amssymb}
\usepackage[hidelinks]{hyperref}
\usepackage{slashed}
\usepackage{graphicx,epsfig}
\usepackage{caption}
\usepackage{xcolor}
\usepackage{slashed,cancel,calrsfs}
\usepackage[12hr]{datetime} 
\hypersetup{colorlinks,linkcolor={blue!50!black},citecolor={blue!50!black},urlcolor={blue!80!black}}

\def\met{\cancel{E}_T}
\def\hpp{h_R^{++}}

\def\mhpp{m_{h_R^{++}}}
\usepackage[a4paper, inner=2.0cm, outer=2.0cm, top=2.0cm,bottom=2.0 cm, binding offset=0 cm]{geometry}

\begin{document}
	\preprint{HRI-RECAPP-2023-04}
	\title{\textbf{Search for a leptophobic doubly charged Higgs in same-sign four-lepton and six-lepton signatures in a left-right symmetric model}}

	\author{Tathagata~Ghosh}\email{tathagataghosh@hri.res.in}
	\author{Rafiqul~Rahaman}\email{rafiqulrahaman@hri.res.in}
	\author{Santosh~Kumar~Rai}\email{skrai@hri.res.in}
	\affiliation{Regional Centre for Accelerator-based Particle Physics, Harish-Chandra Research Institute,\\ A CI of Homi Bhabha National Institute, 
		Chhatnag Road, Jhunsi, Prayagraj 211019, India}
    \begin{abstract}
		We investigate the possibility of multi-lepton (four and six) signatures, including an exotic signature of same-sign four-lepton (SS4L) as signals of pair production of a doubly charged Higgs in the minimal left-right symmetric model, extended with two doublet scalars. The right-handed neutrino masses are generated in this model through a dimension-$5$ lepton-number violating operator allowing the triplet scalar interactions with leptons to become negligibly small.  This leads to interesting six-lepton and SS4L signatures that can be observed at the high-luminosity phase of the Large Hadron Collider (HL-LHC) with almost no background for doubly charged Higgs with mass below 500 GeV.
	\end{abstract}
	\maketitle
	\section{Introduction}\label{sec:intro}
The left-right symmetric model (LRSM)~\cite{Mohapatra:1974hk, Senjanovic:1975rk} is one of the most well-motivated beyond the standard model (BSM) scenarios that aim to address some of the limitations of the standard model (SM). In the LRSM, the electroweak (EW) theory is based on the gauge group $SU(2)_L\otimes SU(2)_R \otimes U(1)_{B-L}$ with $B$ and $L$ representing the Baryon and Lepton numbers, respectively. Here, unlike in the SM,  both the left- and right-handed quarks and leptons have similar representations under the gauge symmetry, where the left-handed particles form doublets under the $SU(2)_L$ and the right-handed particles form doublets under the $SU(2)_R$ gauge group. The anomaly free  $(B-L)$ symmetry is locally gauged in the LRSM replacing the 
arbitrary weak hypercharge $Y$ of the SM by the known quantum numbers $B$ and $L$. The electric charge of a particle is then governed by known quantities as 
\begin{equation}\label{eq:charge-formulae}
Q_{\rm em} = T_{3L} + T_{3R} + \frac{B-L}{2} ,
\end{equation}
where $T_{3L}$ and $T_{3R}$ are the eigenvalues of the $T_3$ generators of $SU(2)_L$ and $ SU(2)_R$ gauge groups, respectively.		
Parity is an exact symmetry in the LRSM demanding the presence of three right-handed neutrinos which are not present in the SM. If the neutrinos are of Majorana type, the right-handed neutrinos help to generate tiny masses for the left-handed neutrinos and their mixing observed in experiments~\cite{SNO:2002tuh,Super-Kamiokande:1998kpq,KamLAND:2002uet,Esteban:2020cvm,NuFit50} through a combination of Type-I and Type-II seesaw mechanisms~\cite{ Minkowski:1977sc, Sawada:1979dis, Glashow:1979nm, Mohapatra:1979ia} without the need of unnaturally small Yukawa couplings. 
Parity violation occurs naturally through the spontaneous breaking of the left-right symmetry at a scale much higher than the EW scale explaining the observed $CP$-violation in the EW sector~\cite{Mohapatra:1974hk,Babu:1989rb}.
In addition, the LRSM can also account for the observed baryon asymmetry of the universe via leptogenesis through the spontaneous breaking of the $(B-L)$ symmetry~\cite{Kuzmin:1980yp,Frere:1992bd,Sarkar:2007er}.

In a typical LRSM, the bi-doublet scalar plays the crucial role of generating the Dirac masses for the leptons and quarks through its Yukawa interactions, while two scalar triplets, one left-handed and one right-handed, provide Majorana masses to the left-handed and right-handed neutrinos, respectively. The bi-doublet scalar is also responsible for tree-level flavor changing neutral current (FCNC) interactions, suppressed by the $SU(2)_R$ breaking scale. 

The LRSM has a significantly richer bosonic sector than the SM and includes right-handed heavy gauge bosons (charged and neutral) as well as extra neutral, doubly, and singly charged Higgs bosons. The doubly charged Higgs' ($h^{++}$) in LRSM decay to a pair of same-sign charged leptons via their Yukawa coupling and to same-sign charged gauge bosons (on-shell or off-shell depending on the mass of doubly charged Higgs)~\cite{Aoki:2011pz}. However, in both cases, the decay of the $h^{++}$ is dependent on whether it belongs to the $SU(2)_L$ ($h^{++}_L$) or $SU(2)_R$ ($h^{++}_R$) triplets. While the leptonic decay of the $h^{++}$'s is directly dependent on their respective Yukawa couplings, the $W^\pm_L W^\pm_L$ decay mode can be easily suppressed for a sub-TeV mass of the $h^{++}_L$ by choosing a very small vacuum expectation value (vev) for the $SU(2)_L$ triplet. At the same time,  the $h^{++}_R$ decay to $W^\pm_L W^\pm_L$ is also absent if $W_L - W_R$ mixing is zero. Our intention in this article is to probe the doubly charged Higgs' in the low mass range ($<1$ TeV). 

Experiments have already ruled out low mass doubly charged Higgs boson in di-leptonic decay mode.  For example, ATLAS Collaboration has ruled out $m_{h^{++}} < 1080$ GeV in searches involving their 
di-leptonic decay~\cite{ATLAS:2022pbd} only, assuming they decay with equal branching fractions in electron ($e$) and muon ($\mu$) channel. On the other hand, for $h^{++}$ decaying to $W^+_L W^+_L$ (or $W^+_L W^{+\star}_L$ if $m_{h^{++}}<2m_{W_L}$) mode with no leptonic decay, ATLAS Collaboration has put an upper limit of $350$ GeV on the mass of $h^{++}$~\cite{ATLAS:2021jol}. 

In order to search for an experimentally allowed doubly-charged Higgs in the sub-TeV range, it is quite clear that its leptonic decay mode must be quite suppressed.
In this work, we, therefore, adopt a framework in which the triplet scalars have tiny Yukawa couplings to the leptons to suppress the leptonic decay of the doubly charged Higgs. Choosing a right-handed doubly charged Higgs as the lighter state offers us an opportunity to probe lower mass $h^{++}$ in the multi-lepton final state signature (four and more leptons) through the four-body decay to right-handed Majorana neutrino and leptons via  $h_R^{++}\to W_R^{+\star} W_R^{+\star}$ decay mode. To achieve this mode as the primary decay channel, one needs to choose  Yukawa couplings ($<10^{-7}$)  to suppress the 
di-lepton decay width compared to the four-body decay width. The right-handed neutrino masses however become quite light, in the range of few MeV for a typical right-handed symmetry breaking scale of $\sim 10$ TeV. 
This interplay between the symmetry breaking scale (viz. the $W_R$ mass) which controls the partial decay width of the doubly-charged scalar into the four-body decay mode and the heavy neutrino mass (viz. the Yukawa coupling) which controls its di-lepton 
decay width, makes it challenging to simultaneously generate the light neutrino masses
 as well as make the four-body decay mode dominant. For example, a few $100$ GeV mass of right-handed neutrinos, which can produce neutrino masses in $0.1$ eV range through Type-I seesaw, set the mass of $W_R$ to ${\cal O} (10^{6})$ TeV, which would, in turn, push the Yukawa couplings to much larger values and make the di-lepton channel dominant. For the doubly charged Higgs to remain relatively leptophobic and the $W_R$ mass to be around $10$ TeV, we 
 must invoke an alternate mechanism to generate the 
 right-handed neutrino masses. This can be achieved by introducing a right-handed scalar doublet (along with a left-handed doublet for Parity symmetry) via a dimension-$5$ effective operator. In this scenario, the $W_R$ mass which now has contributions from both the scalar triplet and a scalar doublet can be in a few TeV range, while we can get very small Yukawa couplings of the leptons with the scalar triplets. The origin of these scalar doublets could be from a heavy scale framework that is decoupled from the left-right breaking scale, much like the right-handed triplet
 scalar is decoupled from the breaking of the SM EW scale.

In the above scenario, the doubly charged Higgs in the right-handed sector, despite being leptophobic produces interesting and rare multi-lepton collider signatures through the  four-body decay mode via two off-shell same-sign $W_R$. These signatures include four-lepton, six-lepton, and very interestingly a same-sign four-lepton (SS4L) final states, which could be observed in future experiments as clear channels of new physics signals.

The rest of the article is arranged as follows. In section~\ref{sec:model}, we briefly describe our model in consideration. We discuss our multi-lepton signals and possible SM backgrounds at the LHC in section~\ref{sec:sigbkg}. Section~\ref{sec:result} contains the findings of our analysis followed by prospects of the multi-lepton final sates in future lepton colliders in section~\ref{sec:lep-col}. We conclude in section~\ref{sec:conclusion}.

\section{Model description}\label{sec:model}
    In this section, we present a brief overview of our model. We consider a minimal LRSM with the Higgs sector further extended by two Higgs doublets. An LRSM is invariant under the gauge group  $SU(3)_c\otimes SU(2)_L\otimes SU(2)_R \otimes U(1)_{B-L}$ with $B$ and $L$ being the Baryon and Lepton numbers, respectively. The quarks and leptons transform under the gauge symmetry as~\cite{Deshpande:1990ip}     
	\begin{align}
		Q_L &= \begin{pmatrix}
			u_L \\ d_L
		\end{pmatrix} \in (\mathbf{3},\mathbf{2},\mathbf{1},1/3)\,, & Q_R &= \begin{pmatrix}
			u_R \\ d_R
		\end{pmatrix} \in (\mathbf{3},\mathbf{1},\mathbf{2},1/3)\,, \\
		L_L &= \begin{pmatrix}
			\nu_L \\ \ell_L
		\end{pmatrix} \in (\mathbf{1},\mathbf{2},\mathbf{1},-1)\,, & L_R &= \begin{pmatrix}
			\nu_R \\ \ell_R
		\end{pmatrix} \in (\mathbf{1},\mathbf{1},\mathbf{2},-1)\,,
	\end{align}
     where the numbers within parenthesises represent the $SU(3)_c, \, SU(2)_L, \,  SU(2)_R$ and $ U(1)_{B-L}$ charges of quarks and leptons. Furthermore, the requirement that parity is a symmetry of the model demands $\Psi_L \leftrightarrow \Psi_R$, with $\Psi = Q, L$. The electric charge of particles in an LRSM is given by
     \begin{align}
		Q_{\rm em} &= T_{3L} + T_{3R} + \frac{B-L}{2}\, ,
	\end{align}
	where $T_{3}$ is the third component of the isospin of the fields in a given multiplet.

	The scalar sector of the model consists of a bi-doublet
	\begin{align}
		\Phi &=\begin{pmatrix}
			\phi^{0}_{1}   &   \phi^{+}_{1} \\
			\phi^{-}_{2}   &   \phi^{0}_{2}
		\end{pmatrix} \in  (\mathbf{1},\mathbf{2},\mathbf{2},0) \,, 
	\end{align}		
	two triplets
	\begin{align}
		\Delta_{L} &=\begin{pmatrix}
			\frac{\delta^{+}_{L}}{\sqrt{2}}& \delta^{++}_{L} \\
			\delta^{0}_{L}                 & -\frac{\delta^{+}_{L}}{\sqrt{2}}
		\end{pmatrix}\in (\mathbf{1},\mathbf{3},\mathbf{1},2)\,, 
		& \Delta_{R} &=\begin{pmatrix}
			\frac{\delta^{+}_{R}}{\sqrt{2}}& \delta^{++}_{R} \\
			\delta^{0}_{R}                 & -\frac{\delta^{+}_{R}}{\sqrt{2}}
		\end{pmatrix}\in (\mathbf{1},\mathbf{1},\mathbf{3},2)\,,
	\end{align}
	and two doublets~\cite{Chakrabortty:2010zk} 
 \begin{align}
		H_L &= \begin{pmatrix}
			h_L^+ \\ h_L^0
		\end{pmatrix} \in (\mathbf{1},\mathbf{2},\mathbf{1},1)\,, & 
		H_R &= \begin{pmatrix}
			h_R^+ \\ h_R^0
		\end{pmatrix} \in (\mathbf{1},\mathbf{1},\mathbf{2},1)\,. 		
	\end{align}
The vacuum structure with the scalar vev's which contribute to the different stages of symmetry breaking are given by
\begin{equation}\label{eq:scalar-vacumm}
\langle \Phi \rangle = \frac{1}{\sqrt{2}} \begin{pmatrix} v_1 & 0 \\ 0 & v_2  \end{pmatrix},~~
\langle \Delta_{L(R)} \rangle = \frac{1}{\sqrt{2}}\begin{pmatrix} 0 & 0 \\ v_{tL}(v_{tR}) & 0  \end{pmatrix},~~
\langle H_{L(R)} \rangle = \frac{1}{\sqrt{2}}\begin{pmatrix} 0  \\ v_{L}(v_{R})   \end{pmatrix},
\end{equation}
where $v_2$ can be complex in general. The EW vacuum vev will be given by
$v_{\rm EW} = \sqrt{v_1^2+v_2^2+v_{tL}^2+v_L^2}=246$ GeV. In our model, we choose $v_2=0$,  $v_{L}\simeq0$, and $v_{tL}\simeq 0$, while $v_{tR}$ and $v_R$ set the scale where  $SU(2)_R\otimes U(1)_{B-L}$ is broken to $U(1)_Y$.
The Lagrangian for the Higgs sector in this model is given by 	
	\begin{eqnarray}\label{eq:Lhiggs}
		{\cal L}_{{\rm Higgs}} &=& {\rm Tr}[(D_\mu\Phi)^\dagger(D^\mu\Phi)] 
		+ {\rm Tr}[(D_\mu\Delta_L)^\dagger(D^\mu\Delta_L)] +
		{\rm Tr}[(D_\mu\Delta_R)^\dagger(D^\mu\Delta_R)] \nonumber\\
		&+&(D_\mu H_L)^\dagger(D^\mu H_L) +(D_\mu H_R)^\dagger(D^\mu H_R) 
		+ V(\Phi,\Delta_{L,R},H_{L,R}) ,
	\end{eqnarray}	
	where
	\begin{eqnarray}
		D_\mu	\Delta_{L,R} &=& \partial_\mu \Delta_{L,R} 
		- i g  \left[ \frac{1}{2} W_{L,R\mu}^a\sigma^a ,\Delta_{L,R}  \right] - i g_{BL} B_\mu\Delta_{L,R}, \nonumber\\
		D_\mu\Phi&=& \partial_\mu\Phi   - \frac{i g}{2} \left[ W_{L\mu}^a\sigma^a\Phi - \Phi W_{R\mu}^a\sigma^a \right] ,\nonumber\\
		D_\mu H_{L,R} &=& \partial_\mu H_{L,R}  - \frac{i g}{2} W_{L,R\mu}^a\sigma^a H_{L,R} - i \frac{1}{2}g_{BL} B_\mu H_{L,R} ,
	\end{eqnarray}
 are the covariant derivatives for the scalar fields. Both $SU(2)_L$ and $SU(2)_R$ gauge couplings are given by a common coupling $g \equiv g_L = g_R$ following the left-right symmetry of the model. On the other hand, the notation $g_{BL}$ is used for the $U(1)_{B-L}$ gauge coupling in the above equations. Due to the rich scalar sector of the model, the scalar potential, $V(\Phi,\Delta_{L,R},H_{L,R})$ is quite involved and contains many terms. So, for brevity we present $V(\Phi,\Delta_{L,R},H_{L,R})$ with all the renormalizable terms in Eq.~(\ref{eq:scalar-potential}) of appendix~\ref{sec:potential}. 
	The kinetic terms in Eq.~(\ref{eq:Lhiggs})  generate the masses for the neutral and charged gauge bosons as
	\begin{eqnarray}
		m_{Z_\mp}^2 &\simeq& \frac{1}{8} \Bigg(
		g^2 v_1^2+g_{BL}^2 \left(v_R^2+4 v_{tR}^2\right)+g^2 \left(v_1^2+v_R^2+4 v_{tR}^2\right)\nonumber\\
		&\mp&\bigg[\left(g^2 v_1^2+g_{BL}^2 \left(v_R^2+4 v_{tR}^2\right)+g^2 \left(v_1^2+v_R^2+4 v_{tR}^2\right)\right)^2\nonumber\\
		&-& 4 v_1^2 \left(v_R^2+4 v_{tR}^2\right) \left(g^2 \left(g_{BL}^2+g^2\right)+g_{BL}^2 g^2\right)\bigg]^{1/2} \Bigg),
	\end{eqnarray}
	and 
	\begin{eqnarray}
		m_W&=& \frac{1}{2}g v_1,\nonumber\\
		m_{W_R} &\simeq& \frac{1}{2}g\sqrt{v_1^2+v_R^2+2 v_{tR}^2} 
	\end{eqnarray}
after the scalars get their vevs as given in Eq.~(\ref{eq:scalar-vacumm}). We denote the SM gauge boson $Z=Z_-$ and the heavy neutral right-handed gauge boson as $Z_R = Z_+$ in the above equation.

	The physical scalar spectrum consists of six $CP$-even neutral states with the lightest one being the SM Higgs, four $CP$-odd pseudo scalars, four singly charged Higgs, and two doubly charged Higgs.  
From the scalar potential given in Eq.~(\ref{eq:scalar-potential}) one can evaluate the masses of the doubly charged Higgs bosons as 
\begin{eqnarray}\label{eq:mass-hpp}
 m_{h_{L,R}^{\pm\pm}}^2 &=& \frac{1}{4}\left[ 2\alpha_3 v_1^2 + (3\eta_H^\prime+\eta_{H_{RL}}-\eta_H) v_R^2 -(2\rho_1-4\rho_2-\rho_3)v_{tR}^2 +2\sqrt{2}v_R^2\frac{\xi_H}{v_{tR}} \right.\nonumber\\
 &\pm&\left.\sqrt{\left(4\beta_3^2 v_1^4 + \left(\eta_H^\prime-\eta_{H_{RL}}+\eta_H \right)v_R^2+ (2\rho_1+4\rho_2-\rho_3)v_{tR}^2 \right)}\right].
 \end{eqnarray}
 In the above equation, the plus sign in $\pm$ corresponds to the $h_L^{++}$, while the minus sign  corresponds to the $\hpp$ state.

The masses of physical $CP$-even, $CP$-odd, and singly charged scalars are obtained by numerically diagonalizing the corresponding mass matrices given in appendix~\ref{sec:potential}. 

After the EW symmetry breaking (EWSB), the bi-doublet, $\Phi$,  provides Dirac masses to the SM quarks and leptons via the following Yukawa interaction:	
	\begin{align} \label{eq:model:dirac_masses}
		-\mathcal{L}_Y^\Phi = \overline{Q_L}\left(Y_{Q_1}\Phi +Y_{Q_2} \tilde{\Phi}\right) Q_{R} +
		\overline{L_L}\left(Y_{L_1}\Phi +Y_{L_2} \tilde{\Phi}\right) L_{R} + {\rm H.c.}\,,
	\end{align}
	where $\tilde{\Phi}\equiv -\sigma_{2}\Phi^{\ast}\sigma_{2}$, and we do not specify the flavor indices which is implicit. 	
 Once the $CP$-even components of the bi-doublet acquire vevs, the quark mass matrices in the flavor basis can be written as
	\begin{equation}
		M_U  = \left( Y_{Q_1} v_1 \ + \ Y_{Q_2} v_2^*\right)/\sqrt{2}~~~{\rm and}~~~
		M_D = \left( Y_{Q_1} v_2 \ + \ Y_{Q_2} v_1^*\right)/\sqrt{2},
	\end{equation}
	with vevs $v_1= \sqrt{2}\langle \phi_1^0 \rangle$ and $v_2 =\sqrt{2} \langle \phi_2^0 \rangle$.
	Similarly, the charged lepton mass matrix takes the form,
	\begin{equation}
		M_e= \left(Y_{L_1}  v_2  + Y_{L_2} v_1^*\right)/\sqrt{2}.
	\end{equation}
	Finally, the Dirac mass matrix for the neutrinos is given by
	\begin{equation}\label{eq:dirac-numass}
		m_\nu^D= \left(Y_{L_1} v_1  +  Y_{L_2} v_2^*\right)/\sqrt{2}.
	\end{equation}
	This resembles the scenario where right-handed neutrinos are added to the SM and the tiny Dirac neutrino masses are realized through unnaturally small Yukawa couplings.

 However, in LRSM, the gauge symmetry of the model allows one to write the following lepton number violating Yukawa terms for the two $SU(2)$ triplets:
	\begin{align} \label{eq:model:majorana_masses}
		-\mathcal L_Y^{\Delta} &= \overline{L_L^C} \,Y_{\Delta_{L}}\,(i\sigma_2) \Delta_L\, L_L 
		+\overline{L_R^C} \, Y_{\Delta_{R}}\, (i\sigma_2) \Delta_R\, L_R 
		+ {\rm H.c.}\,.
	\end{align}		
 This naturally leads to Majorana masses ($\sim Y_{\Delta_i} v_{\Delta_i}$) for both the $\nu_L$ and $\nu_R$ once the triplet scalars acquire vev. 
 The neutrinos then obtain their  masses through the Type-I seesaw mechanism from the mass matrix 
	\begin{eqnarray}
		\label{mnu}
		M_\nu =\begin{pmatrix}
			
			0 & m_{\nu}^D 
			\cr
			(m_\nu^D)^{T}&  M_{\nu_R}
		\end{pmatrix}
	\end{eqnarray}
arranged in the basis ($\nu_{L},\nu_R$). This can generate the tiny neutrino masses through the Type-I seesaw mechanism ($M_{\nu_R} >>~ m_D$): 
 \begin{equation}\label{eq:type1-seesaw}
		m_{\nu_{L}}=m_\nu^D M_{\nu_{R}}^{-1} (m_\nu^D)^{T},
	\end{equation}
 where $M_{\nu_{R}} \sim Y_{\Delta_{R}}v_{\Delta_{R}}$ is the Majorana mass of the right-handed neutrinos and $m_\nu^D$ is given in Eq.(\ref{eq:dirac-numass}). As we shall focus on the phenomenology where the right-handed neutrinos are of $\mathcal{O} (100)$ GeV, the typical size of the Yukawa couplings are
 $Y_{\Delta_{R}} \sim \mathcal{O} (10^{-1}-10^{-2})$ for $v_R \sim \mathcal{O}(1-10)$ TeV. 
 However, in our leptophobic framework for the doubly-charged scalars, the $Y_{\Delta_{L,R}}$ are assumed to be negligibly small\footnote{We keep the LR symmetry intact in the couplings and assume $Y_{\Delta_{L}} = Y_{\Delta_{R}} \simeq 0$.} by construction. Thus, we can not generate $\mathcal{O} (100)$ GeV Majorana masses for right-handed neutrinos with $v_R \sim \mathcal{O}(1-10)$ TeV.   
 
 In order to keep the doubly-charged Higgs leptophobic we need to generate the Majorana masses for the right-handed neutrinos in some other way. A notably simple mechanism for the right-handed neutrinos to obtain their Majorana masses would be from the well-known non-renormalizable dimension-$5$ operator~\cite{Weinberg:1979sa},  
	\begin{equation}
		{\cal O}_{WO}^{(5)}=\frac{\eta_{WO}}{\Lambda } \bar{L}_R^c \tilde{H}_R^\dagger \tilde{H}_R L_R, 
	\end{equation}
	where elements of $\eta_{WO}$ (a symmetric complex $3\times 3$ matrix) are  the strength of the 
	non-renormalizable couplings and $\Lambda$ is the cut-off scale which may be around the unification scale. Note that we require the additional scalar doublets, $H_L$ and $H_R$ introduced in our model to write down the effective operator.
	Once the $H_L$ and $H_R$ acquire the vevs, $v_L$ and $v_R$, respectively,
 the operator generates the mass matrix for right-handed neutrinos as 
	\begin{equation}\label{eq:dim5-operator}
		M_{\nu_R} \approx \eta_{WO} \frac{ v_R^2}{2\Lambda}.
	\end{equation}	
	The dimension-$5$ operator~\cite{Abada:2007ux,Bonnet:2012kz} can be realized at tree level in models with $SU(2)$ singlet fermion~\cite{Minkowski:1977sc,Yanagida:1979as,Gell-Mann:1979vob,Mohapatra:1979ia}, triplet scalar~\cite{Magg:1980ut,Schechter:1980gr,Wetterich:1981bx,Lazarides:1980nt,Mohapatra:1980yp,Cheng:1980qt}, and triplet fermion~\cite{Foot:1988aq}.
	 Now assuming $M_{\nu_R} >>~ m_D$, we have the tiny light neutrino masses as shown in 
 Eq.(\ref{eq:type1-seesaw}).

\subsection{Phenomenological benchmark selections}
	
In this paper, we aim to study the prospect of detecting a leptophobic doubly charged Higgs at the LHC in the $h_R^{\pm \pm} \rightarrow W^{\pm^*}_R W^{\pm^*}_R \rightarrow \ell^{\pm} \nu_R \ell^{\pm} \nu_R$ final state with right-handed neutrinos decaying leptonically. To achieve that we vary the right-handed doubly charged Higgs mass in the range of a few hundred GeV to obtain sizable cross-sections at the LHC. The right-handed heavy neutrino masses are kept at $\sim 100$ GeV. We choose our parameters in such a way that all other BSM scalar states, i.e.,  five extra $CP$-even neutral Higgs, four $CP$-odd neutral and charged Higgs bosons, and the left-handed doubly charged Higgs are heavier than 9 TeV. This choice not only ensures that all these states remain out of reach of the LHC but they also satisfy flavor-changing neutral scalar (FCNS) constraints from meson-antimeson mixings.
 	
We have implemented the model in {\tt SARAH}~\cite{Staub:2013tta,Staub:2015kfa} to generate the  Universal FeynRules Output ({\tt UFO})~\cite{Degrande:2011ua} files 
and {\tt SPheno}~\cite{Porod:2003um,Porod:2011nf} to generate the physical mass spectrum.  
We fix the $W_R$ and $Z_R$ masses to be 5.56 and 8.61 TeV, respectively. To obtain the desired mass spectrum we prefer a simple vacuum structure as given below:
	\begin{equation}\label{eq:vacuum_values}
		v_{tR}=10~\text{TeV},~v_{tL}\simeq 0,~v_{R}=9~\text{TeV},~v_{L}\simeq 0,~v_1=246~\text{GeV},~v_2=0.
	\end{equation}	
The mass spectrum, except $h_R^{++}$, is presented in Table~\ref{tab:mass-spectrum} with the values of the scalar potential couplings mentioned in the caption. We must however point out that more complicated choices, with non-zero values of several other parameters in the scalar potential, could also provide viable mass spectrum, which we chose to avoid to keep things simple. The mass of $h_R^{++}$  is varied in the mass range of $[150,500]$ GeV with an interval of $50$ GeV by changing the single parameter $\eta^\prime_{H}$ and shown in Table~\ref{tab:mass-deltaR}.
	
 \begin{table}\caption{\label{tab:mass-spectrum} The mass spectrum of our benchmark points ({\tt BP}) with input parameters: $v_2=0$,$v_{tL}\simeq 0$, $v_L\simeq 0$, $v_1=246$ GeV, $v_{tR}=10$ TeV, $v_R=9$ TeV,
$(\alpha_1,\alpha_2,\beta_1,\beta_3,\lambda_2,\lambda_4,\rho_1,\rho_2,\rho_3,\rho_4,\beta_H,\eta_{H},\xi_{H_{RL}})=0$, 
			$\alpha_3=12.5,~\lambda_1=0.13,~\lambda_3=1.0$, $\alpha_H=0.1,~\lambda_H=4.0,~\eta_{H_{RL}}=1.0,~\xi_H=2\times10^4$. The parameter $\eta_H^\prime$ is varied to fix the mass of $h_R^{\pm\pm}\equiv \delta_R^{\pm\pm}$ given in Table~\ref{tab:mass-deltaR}. }
		\renewcommand{\arraystretch}{1.50}
		\centering	
		\begin{tabular*}{1.0\textwidth}{@{\extracolsep{\fill}}ll@{}}\hline	
			Particle & Mass in GeV  \\ \hline		
			Neutral gauge boson ($Z,~Z_R$) & $91.1887$, $8.61\times 10^3$ \\ \hline
			Charge gauge boson ($W^\pm,W_R^\pm$) & $80.35$, $5.56\times 10^3$ \\ \hline
			$CP$-even scalar ($h,~h_i$) &  $125.54,~(9.27,~9.88,~25.0,~26.01,~33.29)\times 10^3$  \\ 
			$CP$-odd scalar ($A_i^0$) & $(9.88,~25.0,26.08,~33.29)\times 10^3$   \\ \hline
			Singly charged scalar ($h_i^\pm$)  &  $(9.89,~14.08,~25.0,~33.29)\times 10^3$   \\ \hline
			Doubly charged scalar ($h_L^{\pm\pm}\equiv \delta_L^{\pm\pm}$)  & $9.9\times 10^3 $ \\ \hline
			Heavy neutrinos ($\nu_{iR}$) & $50.22,~70.07,~250.29$ \\ \hline
		\end{tabular*}	
	\end{table}
	\begin{table}\caption{\label{tab:mass-deltaR} Mass of $\hpp$ as a function of $\eta_H^\prime$ with all other parameters fixed at their values given in Table~\ref{tab:mass-spectrum}.}
		\begin{tabular}{|c|c|}\hline
			$\eta_H^\prime$ & $\mhpp$ (GeV) \\ \hline
			$-1.41861$   & $150.0$ \\ 
			$-1.41839$  & $201.0$ \\ 
			$-1.41811$ & $250.0$ \\ 
			$-1.4177$ & $300.0$ \\ 
			$-1.41737$ & $350.0$ \\ 
			$-1.41691$   & $400.0$ \\ 
			$-1.41638$ & $450.0$ \\ 
			$-1.4158$  & $500.0$ \\ \hline
		\end{tabular}
	\end{table}
	
	The values of the symmetric Wilson co-efficient matrix elements of the dimension-$5$ operator in Eq.~(\ref{eq:dim5-operator}) are set at
	\begin{equation}
		\frac{\eta_{WO}}{\Lambda}=
		\begin{pmatrix}
			-0.257426 & -1.03972 & 1.93195 \\ 
			& -2.11799 & 0.576741 \\
			&& -1.00915
		\end{pmatrix}\times 10^{-7}
	\end{equation}
	which satisfy the neutrino masses $6.82\times 10^{-5}~\text{eV}^2<\Delta m_{21}^2< 8.04\times 10^{-5}~\text{eV}^2$, 
	$2.43\times 10^{-3}~\text{eV}^2<\Delta m_{31}^2<2.60\times 10^{-3}~\text{eV}^2$  and $U_{PMNS}$  mixing elements within $3\sigma$ error~\cite{Esteban:2020cvm,NuFit50}. One should note that if the cutoff scale $\Lambda$ is around $10$ TeV, elements of $\eta_{WO}$ are ${\cal O}(10^{-3})$.

Next, we focus on the bounds of the model parameters. Below we list several constraints, both theoretical and phenomenological in nature, which are considered while choosing our benchmark points.
	\begin{description}
		\item[Theoretical constraints :] Various theoretical bounds can be imposed on the Higgs sector couplings of the model. We make sure that these couplings do not violate perturbativity and unitarity, and that the scalar potential is bounded from below~\cite{Chakrabortty:2016wkl}.  The perturbativity condition requires that all quartic couplings in the scalar potential are numerically below $4\pi$ at the EW scale~\cite{Mohapatra}. The tree-level unitarity in the scattering of Higgs bosons and longitudinal components of EW gauge bosons demands that eigenvalues of the scattering matrices are numerically below $16\pi$~\cite{PhysRevLett.38.883, PhysRevD.16.1519}. 
   
       \item[The $\rho$ parameter and other EW precision observables:]  The $\rho$ parameter from the EW precision data~\cite{Workman:2022ynf} allows $v_{tL}$ and $v_L$ to be ${\cal O}(1)$ GeV. In our case, for simplicity, we choose  $v_{tL}=v_L\simeq 0$. Then the EW oblique parameters $S$, $T$, and $U$ further constrain the mass differences between the left-handed triplet members to be less than $40$ GeV~\cite{Chun:2012jw,Das:2016bir,Primulando:2019evb,Aoki:2012jj,Dutta:2014dba,Ghosh:2022fzp}. Similar bound applies to the mass-splittings of the neutral and charged components of the left-handed doublet as well~\cite{Jana:2020pxx}. In our case, mass splitting is chosen to be less than $10$ GeV.  In our specific benchmark with $v_2=0$, there is no $W-W_R$ mixing, eliminating the possibility of any contribution to the EW precision observables coming from the large right-handed triplet mass splittings.

\item[Flavour changing neutral scalar:] 
  The bi-doublet in LRSM gives rise to tree-level Higgs-mediated flavor-changing neutral currents (FCNCs) in the quark sector through a non-diagonal CKM matrix. Precisely measured meson-antimeson mixings, such as $K$-$\bar{K}$ mixing and $B_{d,s}$-$\bar{B}_{d,s}$ mixing~\cite{ParticleDataGroup:2016lqr}, puts strong bound on the masses of flavor-changing neutral scalar (FCNS).  The $K$-$\bar{K}$ mixing receives a contribution from the $W_R$ gauge boson through a box diagram, which sets a lower limit on its mass of $m_{W_R} > 4$ TeV~\cite{Zhang:2007da}. Moreover, the tree-level FCNCs mediated by the FCNS (both scalar and pseudo scalar), stemming from the bi-doublet scalar, also contribute to $K$-$\bar{K}$ mixing, limiting its mass to be $>15$ TeV~\cite{Zhang:2007da}. The $B_s$-$\bar{B}_s$ mixing on the other hand puts an even stronger bound of $>25$ TeV on the bi-doublet scalar~\cite{Zhang:2007da}. We have ensured this limit by choosing a large value of $\alpha_3\simeq 4\pi$ and $v_{tR}=10$ TeV.

\item[Higgs-to-diphoton decay :] Since $h_R^{++}$ is light in our particle spectrum, it will have a significant contribution to the 125 GeV Higgs decay to a pair of photons~\cite{Martinez:1989bg,Picek:2012ei,Carena:2012xa,Bambhaniya:2013wza}. The couplings, $\alpha$'s in the scalar potential contribute to the Higgs to diphoton decay (see appendix~\ref{sec:higgs-to-diphoton} for details). The  large value of $\alpha_3$, with  $\alpha_{1,2}=0$ gives a large contribution to the  Higgs to di-photon decay width,  pushing it beyond the experimentally measured di-photon signal strength $\mu_{\gamma\gamma}=1.04_{-0.09}^{+0.10}$~\cite{ATLAS:2022tnm} in the minimal setup of LRSM. However, this additional contribution to the Higgs to di-photon decay rate can be canceled by adding extra fermions (e.g. dark matter)~\cite{Dey:2022whc} in our model.

\item[Resonant heavy gauge boson searches:]  The LHC experiments have set strong bounds on heavy gauge boson masses through di-jet, di-lepton, and heavy neutrino searches. One major constraint comes from the search for $W_R$ in the $\ell^\pm  \nu_R$ decay channel. A recent search~\cite{CMS:2021dzb} by CMS provides the stringiest constraint on the mass of $W_R$ in the $\ell^\pm  \nu_R$ decay channel with the $\nu_R$ decaying to a same-flavor isolated lepton and two jets as well as into a fatjet containing a same-flavor high $p_T$ lepton. The CMS search~\cite{CMS:2021dzb}  places a lower bound on the mass of $W_R$, excluding masses below $4.7$ TeV and $5.0$ TeV for the $e^\pm$ and $\mu^{\pm}$ channels, respectively\footnote{While preparing our manuscript we came across the ATLAS search in Ref.~\cite{ATLAS:2023cjo} which provides a stronger limit of $6.4$ TeV on the mass of $W_R$ gauge boson. However, a $6.4$ TeV $W_R$ mass will not change the results in our multi-lepton analysis as the branching fractions of $\hpp$ to various multi-lepton channels through off-shell $W_R$  will remain the same, as will be discussed later in section~\ref{sec:sigbkg}.}.  
  In the LRSM, the $Z_R$ gauge boson is generally heavier than the $W_R$ gauge boson. Consequently, a stringent constraint on the mass of the $W_R$ boson indirectly imposes a tight constraint on the mass of the $Z_R$ boson, which is approximately 1.5 times the mass of $W_R$.  Multiple searches conducted at the LHC have been aimed at constraining the mass of an additional heavy neutral Higgs boson, whether it be scalar or pseudoscalar, in different production and decay channels~\cite{CMS:2021yci,ATLAS:2022eap,ATLAS:2022rws}. However, when comparing these limits to the limit on the right-handed $W$ boson ($W_R$), they appear relatively weak, ranging from 1 to 2 TeV. It is worth noting that our additional neutral and singly charged scalar particles have masses exceeding $9$ TeV, rendering them safe from any direct searches.
        
	\end{description}

	\section{Signal and background}\label{sec:sigbkg}
 
 The doubly charged Higgs searches at collider experiments usually involve looking at the four-lepton final state where a pair of same-sign leptons are reconstructed to identify the doubly charged scalar. This final state is naturally very clean at the LHC and gives the strongest available limits on its mass. The more challenging search channel where the doubly charged Higgs decays to a pair of same-sign $W$ bosons leads to less stringent mass limits. In this case, both in the multi-lepton and lepton plus jets final states, the reconstruction of the doubly charged scalar is not clean due to missing particles in the final states. In our model, we have another interesting proposition for the doubly charged Higgs search, when it does not couple to the charged leptons or the light SM $W$ bosons. In this section, we present the signal 
 for such a doubly charged Higgs in a host of multi-lepton ($n_\ell \ge 4$)
 channels.

 \subsection{Four-lepton, six-lepton, and eight-lepton signal processes}\label{sec:signalprc}

	The signal process of our interest is the pair production of right-handed doubly charged Higgs bosons at the LHC, i.e.,
	\begin{equation}\label{eq:hr-prod}
		pp\to h_R^{++} h_R^{--}
	\end{equation}
	followed by the four-body decays of each of the doubly charged Higgs through two off-shell $W_R^\pm$ bosons. Note that in our model, this is the only dominant channel of decay for the lighter right-handed doubly charged Higgs.
	 The $W_R$, which can decay to $\ell \nu_R$ and $jj$ modes, allows the following four-body decay modes for the doubly charged Higgs:
	\begin{subequations}\label{eq:decay-fourbody}
		\begin{align}
			\ell^\pm\nu_R2j:~~ 		   h_R^{\pm\pm} &\to {W_R^\pm}_1^\star {W_R^\pm}_2^\star;~ {W_R^\pm}_{1}^\star \to \ell^\pm \nu_R,~{W_R^\pm}_{2}^\star \to jj,\label{eq:decay-lnu2j}	\\
			2\ell^\pm 2\nu_R:~~		h_R^{\pm\pm} &\to	{W_R^\pm}_1^\star {W_R^\pm}_2^\star,~ {W_R^\pm}_{1,2}^\star \to \ell^\pm \nu_R,~\rm{and}\label{eq:decay-2l2nu}\\	   
			4j:~~		h_R^{\pm\pm} &\to	{W_R^\pm}_1^\star {W_R^\pm}_2^\star,~ {W_R^\pm}_{1,2}^\star \to jj.\label{eq:decay-4j}	   		
		\end{align}
	\end{subequations}
	 Eventually, the $\nu_R$, which has a tiny mixing with the SM neutrinos, decay to a charged lepton and a pair of jets through another off-shell $W_R$. Thus, one can get a four-lepton, six-lepton, and even an eight-lepton final state, or a fully hadronic eight jets final state at the end. Note that the $\nu_R$ dominantly decays only in the semi-leptonic channel  as the $\nu_L$-$\nu_R$ mixing is quite small for the lightest two $\nu_R$.
  
	\begin{figure}[t]
		\centering
		\includegraphics[width=0.7\textwidth]{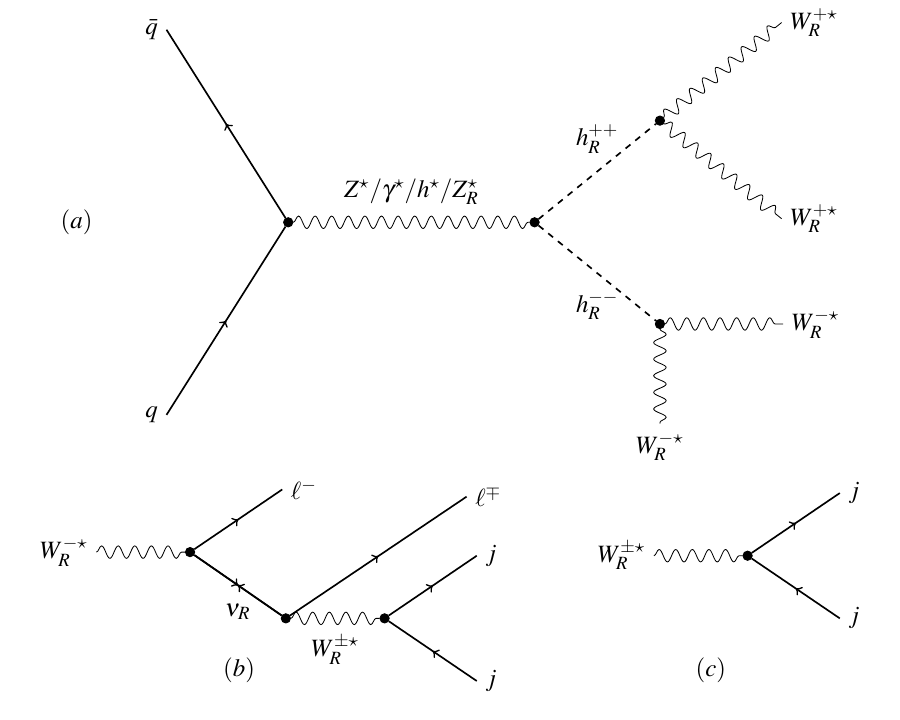}
		\caption{\label{fig:Feynman-diags} Representative Feynman diagrams for the multi-lepton final state in the pair production right-handed doubly charged Higgs.}
	\end{figure}

	The four-lepton final state can be obtained in two ways: when both the $h_R^{\pm\pm}$ decay to $\ell^\pm \nu_R2j$ as shown in Eq.~(\ref{eq:decay-lnu2j}), or when one $h_R^{\pm\pm}$ decays to $2\ell^\pm 2\nu_R$, while the other one decays to $4j$ as shown in Eq.~(\ref{eq:decay-2l2nu}) and Eq.~(\ref{eq:decay-4j}), respectively. 
	The Majorana nature of $\nu_R$ 
 allows its decay to both positive and negative charged lepton, i.e., $\nu_R\to\ell^\pm jj$, thereby giving the interesting possibility where all four charged leptons can be of the same sign,  leading to a rare signal of the same-sign four-lepton (SS4L) final state.
	A six-lepton final state can be obtained when one of the doubly charged-$h_R$ decays to $\ell^\pm \nu_R2j$ while the other one decays to $2\ell^\pm 2\nu_R$. The additional three charged leptons arise when the $\nu_R$ decays in the semi-leptonic channel. In the case of the eight-lepton final state, both the doubly charged-$h_R$ need to decay to $2\ell^\pm 2\nu_R$. As before, the $\nu_R\to\ell^\pm jj$ decay is implicitly assumed in each case.

	The representative Feynman diagrams for these final states are shown in Fig.~\ref{fig:Feynman-diags}. While the diagram in Fig.~\ref{fig:Feynman-diags}$(a)$ shows the pair production of doubly charged Higgs bosons and their decays via off-shell $W_R$s, Figs.~\ref{fig:Feynman-diags}$(b)$ and \ref{fig:Feynman-diags}$(c)$ show the appearance of our desired final states mediated by the off-shell $W_R$s. 
	We label the exclusive $4\ell$,  $6\ell$, and $8\ell$ final states with their intermediate states  $2\ell 2\nu_R4j$, $3\ell 3\nu_R2j$, and $4\ell 4\nu_R$, respectively for later uses. 

 \begin{figure}[t]
		\centering
		\includegraphics[width=0.49\textwidth]{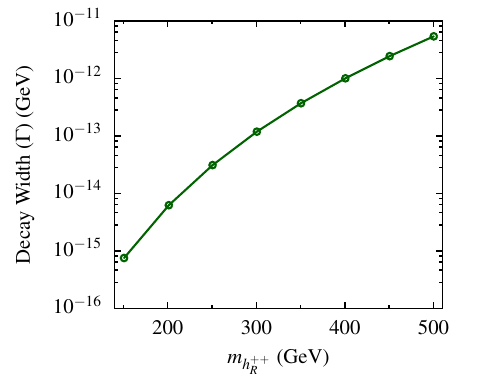}
		\includegraphics[width=0.49\textwidth]{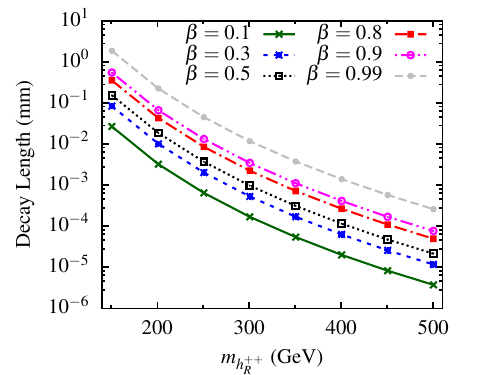}
		\caption{\label{fig:decay-WL} Total decay width ({\em left-panel}) and decay length ({\em right-panel}) for various choices of boost ($\beta$) of $h_R^{++}$ as a function of its mass.}
	\end{figure}
	
 
 We use {\tt MadGraph5\_aMC@NLO}v2.6.7~\cite{Alwall:2014hca} to calculate the four-body decay width and branching ratios of $\hpp$ in the decay modes given in Eq.~(\ref{eq:decay-fourbody}) in the mass range $[150,500]$ GeV, using an interval of $50$ GeV (benchmark points shown in Table~\ref{tab:mass-deltaR}).
 Since the $\hpp$ decays via a very heavy off-shell $W_R$, the decay width is suppressed by the large mass of $W_R$ and can be very small. This also may lead to the  possibility of a very long-lived doubly charged Higgs not decaying within the detector or giving a displaced vertex signal at detectors.  We check such a possibility by estimating its decay length $L$  which for any particle is given by~\cite{Banerjee:2019ktv}
	\begin{equation}\label{eq:decay-length}
		L=\gamma\beta c\tau,
	\end{equation}
	where $\beta$ is the particle's boost, $\gamma=1/\sqrt{1-\beta^2}$ is the relativistic factor, $c$ is the speed of light, and $\tau$ is the decay time of the particle given by $\tau=\hbar/\Gamma$, with $\Gamma$ being the particle's decay width. 
 We evaluate the decay length of the $\hpp$ for a few choices of $\beta$ and show them as a function of $\mhpp$ in the {\em right-panel} of Fig.~\ref{fig:decay-WL}, along with the total decay width in the {\em left-panel} of the same figure.
 One notices that the decay length can be a maximum of a few mm for highly boosted $\hpp$ ($\beta=0.99$) for the lower mass of $150$ GeV. As the mass increases the decay length goes down exponentially. In this work, we work in the limit where  we expect that the $\hpp$ will decay promptly within the detector and will not produce a displaced vertex or leave the detector without decaying~\cite{CMS:2021tkn,Ito:2018asa,CMS:2018qxv,Jana:2020qzn}. 
The decay length remains less than $1$ mm even for heavier $W_R$ (for the details see appendix~\ref{sec:decay-length-fWR}).
 
\begin{figure*}[t]
		\centering
		\includegraphics[width=0.49\textwidth]{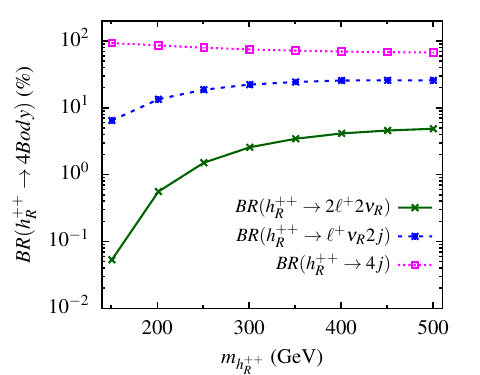}
		\includegraphics[width=0.49\textwidth]{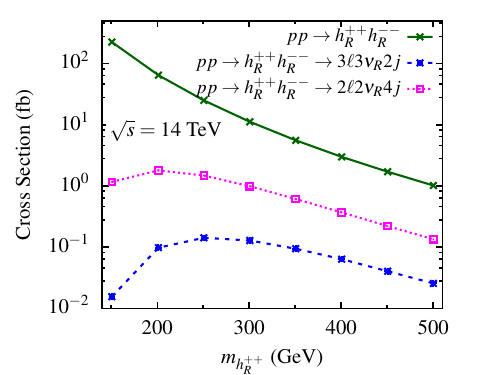}
		\caption{\label{fig:decay-production} Four body decay branching ratios of $h_R^{++}$ to lepton and heavy neutrinos ({\em left-panel}) and the production cross-section of $h_R^{++}h_R^{--}$ scaled to NLO with a naive $k$ factor at $14$ TeV LHC together with their decay to obtain a four-lepton and six-lepton final state ({\em right-panel}) as a function of the mass of $h_R^{++}$. 
		}
	\end{figure*}

 The four-body decay branching ratios of $\hpp$, as given in  Eq.~(\ref{eq:decay-fourbody}), are shown in the {\em left-panel} of Fig.~\ref{fig:decay-production} for our choice of mass range of $\hpp$. Since the $W_R\to \ell \nu_R$ branching ratio is very small compared to the branching ratio of $W_R\to jj$ mode, we find that $Br(\hpp\to2\ell^+2\nu_R)<BR(\hpp\to\ell^+\nu_Rjj)<<BR(\hpp\to4j)$ in the above range of $\mhpp$. The $4j$ branching ratio reduces from $93\%$ to $70\%$ with increasing $\mhpp$ from $150$ GeV to $500$ GeV. While the $\ell\nu_Rjj$ branching ratio increases from $7\%$ to $25\%$, the $2\ell2\nu_R$ branching ratio only goes upto $5\%$ after beginning at less than $1\%$.  The four-body decay branching ratios of $\hpp$ will remain the same for higher $W_R$ mass as no other channel opens up.

	We also calculate the production cross-section of the pair production of doubly charged Higgs bosons at $14$ TeV LHC using {\tt MadGraph5\_aMC@NLO}v2.6.7~\cite{Alwall:2014hca} at the leading order (LO) in QCD in the chosen mass range of $\hpp$. We employ a dynamic choice of factorization scale given by $\sum M_i^T/2$, where $M_i$ is the 
	transverse mass of final state particles. We use {\tt nn23lo1}~\cite{NNPDF:2014otw} for the parton distribution functions (PDFs). We also estimate the cross-section in the four-lepton and six-lepton final states by multiplying the branching ratios to the production cross-section. In the {\em right-panel} of Fig.~\ref{fig:decay-production} we show the production cross-section and the estimated cross-section in four-lepton and six-lepton final states as a function of $\mhpp$  estimated at the next-to-leading order (NLO) in QCD. We use a naive NLO to LO $k$-factor of $1.15$~\cite{Muhlleitner:2003me,Fuks:2019clu} to obtain the NLO cross-sections.
The cross-sections for the four-lepton ($2\ell2\nu_R4j$) and six-lepton ($3\ell3\nu_R2j$) final states first increase until $\mhpp \simeq 250$ GeV due to an increase in the branching ratios of $\ell\nu_Rjj$ and $2\ell2\nu_R$. We then observe the fall in cross-section due to the overall fall of the pair production cross-section of the doubly charged Higgs as the mass increases further. 
	We do not show the cross-sections in a possible eight-lepton final state as the cross-sections are very small due to very low branching in $2\ell2\nu_R$ mode and  will be extremely difficult to probe even at the very high luminosity option of the LHC.

	\subsection{SM backgrounds}\label{sec:bkgproc}
	The SM processes that can provide a four-lepton final state are pair production of top quarks ($t\bar{t}$), pair production of SM bosons $VV+Vh (V=W/Z)$, associated production of $t\bar{t}$ with one ($t\bar{t}V/h$) or two ($t\bar{t}VV$) SM bosons,  triple gauge boson production ($VVV$), and four gauge boson production ($VVVV$).
	The dominant contributions to $4\ell$ final state come from $t\bar{t}$ and $ZZ$ processes. Subdominant backgrounds that contribute to 	$4\ell$ final state are $t\bar{t}Z$, $Zh$, $t\bar{t}h$, $WWZ$, $WZ$, $WZZ$, $ZWWW$, $ZZZZ$ and $t\bar{t}W$. All other backgrounds are reducible.
	
	 The above-mentioned SM subprocesses could all lead to a six-lepton final state with  the most likely sources being $ZZZ$, $ZZZZ$, and $Zh$ processes giving six or more prompt leptons in the final state. We will show in the next section that the six-lepton final state has nearly zero background. 
	
	\section{Result}\label{sec:result}
	In this section, we study the signal of four-lepton, same-sign four-lepton, and six-lepton final states for the $\hpp$ mass lying within the range of $[150,500]$ GeV.
	We generate the signal events for eight benchmark points (as shown in Table~\ref{tab:mass-deltaR}) and the background processes in {\tt MadGraph5\_aMC@NLO}v2.6.7 with up to two extra jets using the $5$ flavor scheme. The event generation is done at the
	leading order (LO) in QCD. The events are generated without cuts on the final state particles, with a dynamic choice of factorization scale given by $\sum M_i^T/2$, where $M_i$ is the 
	transverse mass of the final state particles. The {\tt nn23lo1} PDF sets are used in our simulation. The parton-level
	events are then passed through {\tt PYTHIA8}~\cite{Sjostrand:2014zea} for showering and 
	hadronization. We also implement the MLM matching scheme~\cite{Hoeche:2005vzu,Alwall:2007fs} with {\tt PYTHIA8} to avoid double counting of jets. Finally, a fast detector simulation is performed with 
	{\tt Delphes v3.4.2}~\cite{deFavereau:2013fsa}.
    For all the SM background processes, we use the {\tt MadLoop}~\cite{Hirschi:2011pa} package available within {\tt MadGraph5\_aMC@NLO}~\cite{Hirschi:2011pa} to decay all the heavy unstable particles in their fully leptonic decay channels (except $h$, which is decayed in {\tt PYTHIA8}).
    The NLO cross-sections for the background subprocesses are shown in the 3rd column in Table~\ref{tab:NEvents-SigBKG} taken
	from the 13 TeV LHC results available in Ref.~\cite{Alwall:2014hca} except for the $t\bar{t}$ process for which an NNLO $k$-factor of $1.6$~\cite{Catani:2019hip} is used.
	
	The final state signal that we consider for our analysis contains  a large number of leptons and jets.  Such large multiplicity of particles in the final state can lead to particles getting merged with little separation between them. This will reduce the signal events considerably as we get  less isolated particles at the detector with the standard isolation criteria used in the present CMS and ATLAS detector. We presume that better isolation of leptons will be possible in the future with an upgraded detector for the high-luminosity run of the LHC.
    A lepton is said to be isolated if the activity in its vicinity is small enough within a cone of radius $\Delta R_{\ell}= \sqrt{\Delta \eta^2 + \Delta \phi^2}$ around it. 
    We use a relatively small isolation cone of $\Delta R_{\ell}^{\rm max}=0.2$~\cite{ATLAS:2021srw,ATLAS:2019jvq,Apollinari:2017lan,ATLAS:2016ukn} for the leptons in our {\tt Delphes} detector simulation instead of $\Delta R_{\ell}^{\rm max}=0.3$ or $0.4$ which are presently used by the CMS~\cite{CMS:2017yfk,CMS:2020uim} and ATLAS~\cite{ATLAS:2019qmc,ATLAS:2020auj} analyses. The activity around the lepton is parameterized by
     the isolation variable, $I_\ell$, defined as
 \begin{equation}\label{eq:lep-isolation}
I_\ell = \frac{ \sum\limits_{i\ne \ell}^{\Delta R < \Delta R_{\ell}^{\rm max},~p_T(i)>p_T^{\rm min}} p_T(i)}{p_T(\ell)}.
\end{equation}
We use $I_{\ell}$ values less than $0.12$ and $0.25$ for electrons and muons, respectively, with 
$p_T^{\rm min}=0.5$ GeV in {\tt Delphes}.
	Finally, the events are selected after the detector simulation with the following requirements on the leptons~\cite{CMS:2017dzg}, 
	\begin{equation}\label{eq:sel-cut}
		p_T(\ell)>10~\text{GeV},~|\eta_e| < 2.5,~|\eta_\mu| < 2.4.
	\end{equation}
	 In order to put our analysis with a loose isolation criterion in perspective with the slightly stronger requirements in current analyses at LHC, we have shown a comparison for one benchmark value of $\mhpp$  with both choices of $\Delta R_\ell^{\rm max}=0.2$ and $\Delta R_\ell^{\rm max}=0.3$ for the lepton isolation criteria in the latter part of our analysis.
	
	\begin{figure*}
		\centering
		\includegraphics[width=0.49\textwidth]{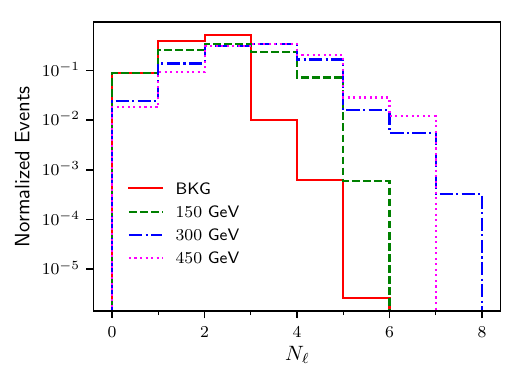}	
		\caption{\label{fig:dist-nlj-full} Normalized distributions of lepton multiplicities for the signal as well as the combined background for $\mhpp$ values of $150$ GeV, $300$ GeV, and $450$ GeV. The leptons satisfy the criteria of Eq.~\ref{eq:sel-cut}.}
	\end{figure*}
	\begin{figure*}
		\centering
		\includegraphics[width=0.49\textwidth]{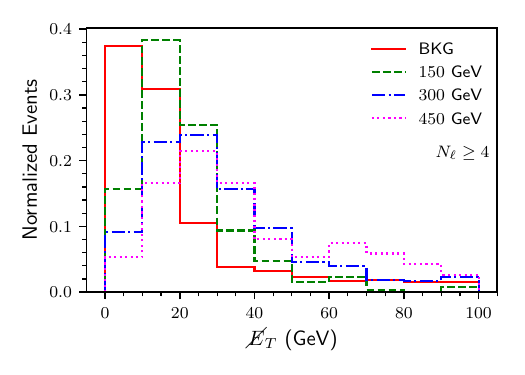}	
		\includegraphics[width=0.49\textwidth]{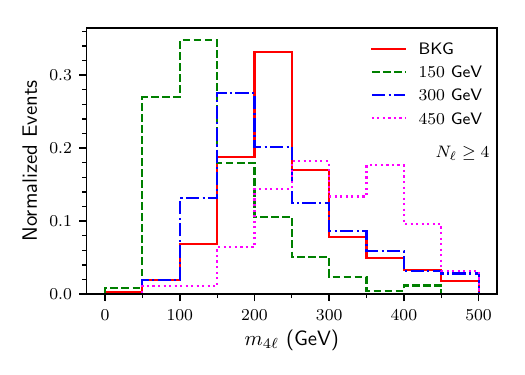}		
		\caption{\label{fig:dist-nj-kin-4lcut} Normalized distributions for  $\met$ and  $m_{4\ell}$ for $N_\ell\ge 4$ (satisfying the selection criteria of Eq.~\ref{eq:sel-cut}) in the signal as well as the combined background for the  $\mhpp$ of $150$ GeV, $300$ GeV, and $450$ GeV.}
	\end{figure*}

    After setting up the criteria for selecting isolated leptons we look at the charged lepton multiplicity,  missing transverse energy ($\cancel{E}_T$), and the invariant mass of four leptons ($m_{4\ell}$).
    Figure~\ref{fig:dist-nlj-full} shows the normalized distributions for the  charged lepton multiplicity for three choices of  benchmark points with $\hpp$ mass of $150$ GeV, $300$ GeV, and $450$ GeV, along with the combined SM background. We combine the $2\ell 2\nu_R4j$ and $3\ell 3\nu_R2j$ channels for the signal benchmark points in this plot. As the heavier $\hpp$ has increased branching probabilities to the leptonic four-body decay modes, we see that the charged lepton multiplicity is higher for the heavier states. In addition, more charged leptons will achieve the threshold energies for detection when they originate from a heavier parent particle, which in this case is the $\hpp$.
	The normalized distributions for $\cancel{E}_T$ and $m_{4\ell}$ are shown in Fig.~\ref{fig:dist-nj-kin-4lcut} after demanding at least four leptons. Similar to the charged lepton multiplicity, $\cancel{E}_T$ increases as $\mhpp$ increases. It is needless to say that for the signals, the peaks in the $m{4\ell}$ distributions arise due to $\hpp$ decays. 
 However, the peaks are shifted towards lower energy than the mass of $\hpp$  as there are extra jets and missing energy in the $4\ell$ final state coming from three different channels, where one of them has no $\hpp$ resonance (cf. Eq.~(\ref{eq:decay-2l2nu})). For the combined SM backgrounds, the peak is coming from the $ZZ$ process, which is the most dominant background as can be seen from Table~\ref{tab:NEvents-SigBKG}.  
 The $ZZ$ process being a $t$-channel one as opposed to a $s$-channel one, the $m_{4\ell}$  peak appears above the threshold energy ($2m_Z$).

	\begin{figure*}
		\centering
		\includegraphics[width=0.49\textwidth]{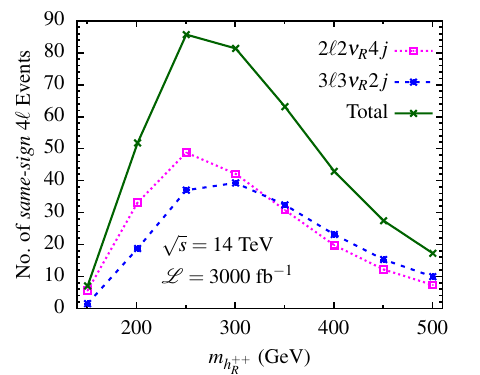}
		\includegraphics[width=0.49\textwidth]{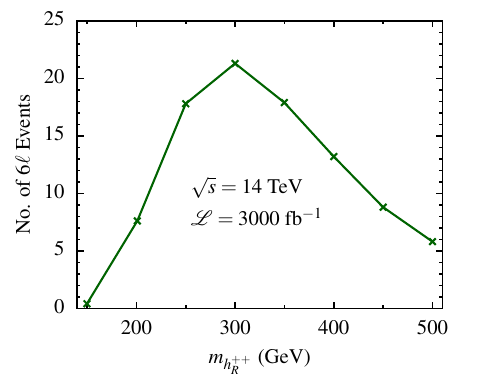}	
		\includegraphics[width=0.49\textwidth]{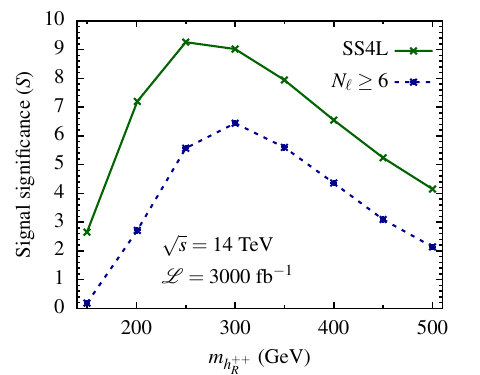}		
		\caption{\label{fig:ss4l6lEvents}The estimated number of same-sign four-lepton (SS4L) ({\em top-left-panel}) and six-lepton events ({\em top-right-panel}) are shown as a function of the mass of  $h_R^{++}$ at $\sqrt{s}=14$ TeV LHC with integrated luminosity of ${\cal L}=3000$ fb$^{-1}$ after selection  cuts at the detector level. The {\em bottom panel} shows the expected signal significance for the SS4L (using Eq.~(\ref{eq:sig-ss4l})) and six-lepton final state (using Eq.~(\ref{eq:sig-6l})).}
	\end{figure*}
 	The most striking signals in our analysis would be the rare SS4L~\cite{Chun:2012zu,Chun:2019hce,Bai:2021ony} with practically no SM background and the six-lepton final state~\cite{Barger:2009xg,Huitu:2017vye,Das:2022cmv} with very low background. 
	We present our findings for the SS4L ({\em top-left panel}) and six-lepton ({\em top-right panel}) events in Fig.~\ref{fig:ss4l6lEvents} projected to the integrated luminosity of ${\cal L}=3000$ fb$^{-1}$. The SS4L events not only come from the $2\ell 2\nu_{R}$ process but also from the $3\ell 3\nu_{R}$ process. In the $2\ell 2\nu_{R}$ process, SS4L events originate only from the configuration which includes  the fully leptonic decay channel of either $h_R^{++}$ or $h_R^{--}$. Though only $8$ SS4L events are expected for $\mhpp=150$ GeV, a maximum of $90$ such events can be observed near $\mhpp=250$ GeV because of the nature of cross-section in this region (see Fig.~\ref{fig:decay-production} for details).
	In the case of the six-lepton final state, the benchmark with $\mhpp=150$ GeV is not a viable option, but $6$ to $21$ such events can be observed for $\mhpp$ in the range of $200$-$500$ GeV.

	\begin{small}
		\begin{table}\caption{\label{tab:NEvents-SigBKG} Generation level cross-sections in fb (scaled to NLO), the number of four-lepton and six-lepton events along with detector efficiencies for various SM backgrounds (BKG) at the $14$ TeV LHC estimated at an integrated luminosity of ${\cal L}=3000$ fb$^{-1}$ after using the selection cuts given in Eq.~(\ref{eq:sel-cut}). 
				All unstable particles are decayed fully leptonically at the generation level, except for the Higgs ($h$), which is decayed in {\tt PYTHIA8}. In the second row, $\epsilon$ stands for efficiency for the lepton multiplicity shown in the first }
			\renewcommand{\arraystretch}{1.20}
			\centering	
			\begin{tabular*}{1.0\textwidth}{@{\extracolsep{\fill}}|llc|cc|cc|@{}}\hline	
				BKG  & Process          &     $\sigma$ (fb)     &\multicolumn{2}{c|}{$N_{\ell}\ge 4$}    &\multicolumn{2}{c|}{$N_{\ell}\ge 6$}    \\\hline
				&&& $\epsilon$ & \#Events& $\epsilon$ &\#Events \\ \hline        
				BKG-1& $t\bar{t}$      &$47256.6747 $&$3.06\times 10^{-5}$& $3975.6 $&$< 7\times10^{-10}$& $0   $\\\hline
				BKG-2& $t\bar{t}Z$     &$3.4625     $&$2.99\times 10^{-1}$& $3110.4  $&$3.29\times 10^{-5}$& $0.3 $\\\hline
				BKG-3& $t\bar{t}W^\pm$  &$6.5974     $&$3.62\times 10^{-3}$& $71.6    $&$<5\times 10^{-6}$& $0   $\\\hline
				BKG-4& $W^+W^-Z$        &$0.5556     $&$2.49\times 10^{-1}$& $415.8   $&$<6\times 10^{-5}$& $0   $\\\hline
				BKG-5& $W^\pm ZZ $      &$0.0590     $&$5.17\times 10^{-1}$& $91.4    $&$< 5.6\times 10^{-4}$& $0 $\\\hline
				BKG-6& $ZZZ $           &$0.0044     $&$6.97\times 10^{-1}$& $9.1     $&$1.46\times 10^{-1}$& $1.9   $\\\hline
				BKG-7& $t\bar{t}ZZ $    &$0.0006     $&$7.83\times 10^{-1}$& $1.4     $&$1.72\times 10^{-1}$& $0.3 $\\\hline
				BKG-8& $t\bar{t}W^\pm Z$&$0.0032     $&$5.52\times 10^{-1}$& $5.3     $&$< 10^{-2}$& $0   $\\\hline
				BKG-9& $W^+W^-ZZ $      &$0.0002     $&$7.79\times 10^{-1}$& $0.4     $&$1.81\times 10^{-1}$& $0.1 $\\\hline
				BKG-10&$t\bar{t}W^+W^-$ &$0.0013     $&$2.81\times 10^{-1}$& $1.1     $&$<2.5\times 10^{-2}$& $0   $\\\hline
				BKG-11& $ZWWW$          &$0.0449     $&$5.77\times 10^{-1}$& $77.6    $&$<7.4\times 10^{-4}$& $0 $\\\hline
				BKG-12& $W^\pm Z$       &$673.6155  $&$5.10\times 10^{-5}$& $103.0    $&$<4.9\times 10^{-8}$& $0   $\\\hline
				BKG-13& $ZZ$            &$62.7140   $&$3.30\times 10^{-1}$& $6214    $&$<5.3\times 10^{-7}$& $0   $\\\hline
				BKG-14& $W^\pm h$       &$322.7483  $&$5.99\times 10^{-5}$& $58       $&$< 10^{-7}$& $0   $\\\hline
				BKG-15& $Zh$            &$52.3821   $&$4.06\times 10^{-3}$& $638    $&$1.50\times 10^{-5}$& $2.4 $\\\hline
				BKG-16& $t\bar{t}h$     &$30.4883   $&$5.88\times 10^{-3}$& $538.1    $&$< 10^{-6}$& $0   $\\\hline
				
				\multicolumn{3}{|c|}{Total BKG}                          & & $ 71236.8$           & & $5.1 $      \\\hline              
			\end{tabular*}	
		\end{table}	
	\end{small}	
	%
	

 Next, we analyze the SM backgrounds mentioned in Section~\ref{sec:bkgproc} for both the SS4L and six-lepton final states.  The generation level background cross-sections in fully leptonic final states are listed in the third column of Table~\ref{tab:NEvents-SigBKG}. We should emphasize again that the generation level cuts are not applied to the final state particles coming from heavy SM particle decays.  
To understand the SS4L background, we first study the four-lepton backgrounds without the same-sign criteria of the leptons.
 We show the number of four-lepton and six-lepton events assuming ${\cal L}=3000$ fb$^{-1}$ in the fifth and seventh columns of the same table. We also present the event selection efficiencies in the fourth and sixth columns for four-lepton and six-lepton events, respectively with the cuts given in Eq.~(\ref{eq:sel-cut}). The background for four-lepton analysis can be further reduced by imposing $b$-veto to suppress $t\bar{t}$, and vetoing on the pair of OSSF $m_{\ell^+ \ell^-}$ around the $Z$-mass to kill the $ZZ$ background. The signal significance in the four-lepton final state can be further improved by a strong cut on $m_{4\ell}$, and by asking for extra jets in the final state and using a cut on the scalar sum of hadronic $p_T$ ($H_T$).

It is clear from Table~\ref{tab:NEvents-SigBKG} that in the six-lepton final state, we have limited sources of SM background that come from the $ZZZ$ and $Zh$ sub-processes, which contribute only a few events at most. Collectively, a total of $\sim 5$ background events are expected in the six-lepton final state with $3000$ fb$^{-1}$ of luminosity. We estimate the signal significance for this channel, and they are shown in the {\em bottom-panel} of Fig.~\ref{fig:ss4l6lEvents} for all the benchmark points with {\em dashed blue} line. 
 In this channel,  the signal significance is calculated using the Asimov formula~\cite{Li:1983fv,Cousins:2007yta,Cowan:2010js}
	\begin{equation}\label{eq:sig-6l}
 S = \sqrt{2\left[(s+b) \log\left(1+\frac{s}{b}\right)-s\right]},
	\end{equation}
 where $s$ and $b$  stand for the total number of signal and background events surviving after all the cuts for a given integrated luminosity.
	As stated earlier, the $150$ GeV mass is not a viable option. However, doubly charged Higgs in the mass range of $250$-$400$ GeV gives enough signal events over the SM background to be observed at $5\sigma$ significance.  
	On the other hand, in the SS4L final state, eventually, no background survives. This can be established by folding in the charge misidentification efficiencies for the charged leptons (in the four lepton events given in the fifth column of Table~\ref{tab:NEvents-SigBKG}), which is of the order of $0.001$ for low $p_T$ electrons~\cite{ATLAS:2019dpa} and ${\cal O}(10^{-5})$ for muons~\cite{CMS:2009fdy,CMS:2019ied}. Thus we find that this SS4L final state can be a primary discovery channel for our model. Since the Asimov formula blows up for $b=0$, we estimate the signal significance for this channel using the formula,
 \begin{equation}\label{eq:sig-ss4l}
S=\frac{s}{\sqrt{s+b}}.
 \end{equation}
   As there are no backgrounds, the signal significance for the SS4L case goes as $\sqrt{s}$, which is shown in {\em bottom-panel} of Fig.~\ref{fig:ss4l6lEvents} with {\em solid green} line for all the benchmark points.

	\begin{table}
		\caption{\label{tab:comp-lep-isolation} Number of  SS4L  and  $6\ell$ events for $\mhpp=250$ GeV for two different values  of $\Delta R_\ell^{\rm max}$  for lepton isolation criteria.}	
		\renewcommand{\arraystretch}{1.2}
		\centering	
		\begin{tabular*}{1.0\textwidth}{@{\extracolsep{\fill}}ccc@{}}\hline	
		 $\Delta R_\ell^{\rm max}$ &  SS4L events & $N_\ell\ge 6$ events \\\hline
			$0.2$ & $85.7$ & $17.8$ \\\hline
			$0.3$ &  $69.6$ & $7.6$\\\hline
		\end{tabular*}
	\end{table}
	Finally, we compare the effects of choosing two different criteria for lepton isolation, viz. $\Delta R_\ell^{\rm max}=0.2$ used in our analysis  with $\Delta R_\ell^{\rm max}=0.3$.  We have re-analyzed our signal for SS4L and six-lepton final states 
 and show the comparison in Table~\ref{tab:comp-lep-isolation} 
 for a single benchmark point with $\mhpp=250$ GeV as an example. The effect of the two different choices for $\Delta R_\ell^{\rm max}$ is more pronounced in the events with larger lepton multiplicity. The SS4L events get suppressed by a factor of $1.2$ with $\Delta R_\ell^{\rm max}=0.3$, while the six-lepton events reduce by a factor of $2.3$. We, however, note that the SM background also reduces in the six-lepton case, keeping the signal significance roughly the same. 
	
	\section{Prospect at future lepton colliders}\label{sec:lep-col}
	\begin{figure*}[ht]
		\centering
		\includegraphics[width=0.45\textwidth]{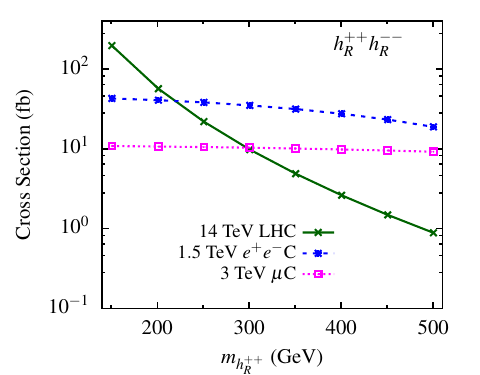}
		\includegraphics[width=0.45\textwidth]{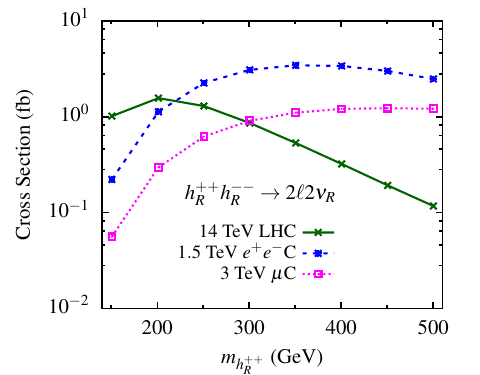}
		\includegraphics[width=0.45\textwidth]{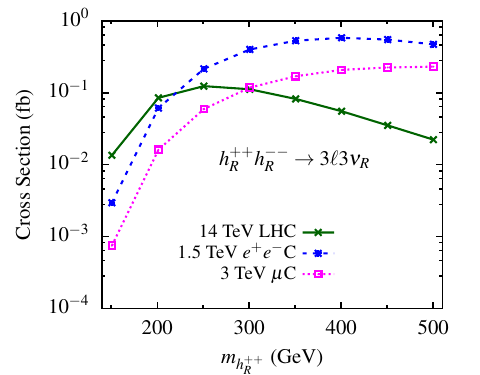}
		\caption{\label{fig:cs-lhc-$e^+e^-$C-muc} The production cross-section of $h_R^{++}h_R^{--}$ (top-left) together with their respective decay branching fractions to obtain the four-lepton ({\em top-right}) and six-lepton ({\em bottom}) final states at the $14$ TeV  LHC, $1.5$ TeV $e^+e^-$C and $3$ TeV muon colliders are shown as a function of the mass of $h_R^{++}$.}
	\end{figure*}
	In this section, we discuss the prospect of observing the four-lepton and six-lepton final states at future lepton colliders, such as an electron-position collider ($e^+e^-$C)~\cite{Djouadi:2007ik,Baer:2013cma,Behnke:2013xla,Bambade:2019fyw,Blaising:2012tz,Brunner:2022usy} and muon collider ($\mu$C)~\cite{Ankenbrandt:1999cta,muon:collmeet,Aime:2022flm,MuonCollider:2022xlm}, and compare them with the LHC case. For the $e^+e^-$C, we consider $\sqrt{s}=1.5$ TeV, while for the muon collider, $\sqrt{s}=3$ TeV is chosen. A comparison plot is shown in Fig.~\ref{fig:cs-lhc-$e^+e^-$C-muc} for the production cross-section of $h_R^{++}h_R^{--}$ ({\em top-left-panel}) along with their decay to a $2\ell 2\nu_R$ ({\em top-right-panel}) and $3\ell 3\nu_R$ ({\em bottom-panel}) final states, which eventually gives $4\ell$ and $6\ell$ final states, respectively. 
	Unlike the $14$ TeV LHC, the production cross-sections at the fixed energy colliders of $e^+e^-$C and $\mu$C do not fall rapidly as the mass of $h_R^{++}$ increases. The production cross-sections are found to be greater for $m_{h_R^{++}}>200$ GeV at the $e^+e^-$C compared to the LHC, while for the $\mu$C, where the center-of-mass energy is higher, the production cross-section exceeds that of the LHC for $m_{h_R^{++}}>300$ GeV. However, the SM backgrounds at the lepton collider are very small and much easier to suppress. For example, the $ZZZ$ background at $e^+e^-$C with $\sqrt{s}=1.5$ TeV is smaller by a factor of 20 than the same at the LHC for the $6\ell$ final state. Thus the $e^+e^-$C and $\mu$C will have a significantly improved performance even for lighter $h_R^{++}$ ($m_{h_R^{++}}<300$ GeV)  compared to the LHC.  For the higher masses, however, the $e^+e^-$C ($>200$ GeV) and $\mu$C ($>300$ GeV) performance will get much better  as the LHC sensitivity begins to drop for heavier masses of $\hpp$ and it will be difficult to observe the $4\ell$ and $6\ell$ final state at LHC beyond a mass of about $450$ GeV. The $e^+e^-$C and $\mu$C  on the other hand will be able to observe the $\hpp$ for much heavier mass, limited only by the energy reach. For example, the $e^+e^-$C cross-section is $18.7$ times larger than the LHC for $m_{h_R^{++}}=500$ GeV, and for the $\mu$C, it is $10$ times larger than the LHC. 
	With a cleaner environment in a lepton collider ($e^+e^-$C and $\mu$C), the number of events in the $4\ell$ and $6\ell$ final states will be even more than the LHC case with a similar amount of integrated luminosity.

\section{Conclusion}\label{sec:conclusion}
    In this study, we investigate the prospect of detecting a doubly charged Higgs boson with lepto-phobic interactions through rare collider signatures of same-sign four-lepton and six-lepton final states. The specific doubly charged Higgs that we are interested in belongs to the right-handed sector of a left-right symmetric model with electroweak gauge structure $SU(2)_L\otimes SU(2)_R\otimes U(1)_{B-L}$, which incorporates a bi-doublet scalar, two triplet scalars, and two doublet scalars. In this setup, we generate the right-handed neutrino masses from the right-handed doublet scalar via a lepton number violating dimension-$5$ effective operator instead of the conventional approach where the right-handed neutrino masses are generated via the leptonic Yukawa interactions of the $SU(2)_R$ triplet. The tiny masses for the left-handed neutrinos are then generated via the type-I seesaw mechanism. An interesting feature emerges in this setup for the doubly charged Higgs which can be made fermiophobic at the LO. As the doubly charged Higgses of $SU(2)_L$ and $SU(2)_R$ do not mix, the light $\hpp$ can only decay via the off-shell exchange of the heavier bosons. In our case, this decay happens through the heavy $W_R$ gauge boson. The setup also helps us to evade the strong experimental bounds on the low masses of doubly charged Higgs boson, coming from the signal analysis of same-sign dilepton and $W^+W^+$ decay channels, and simultaneously keep the right-handed neutrino and right-handed gauge boson masses within the reach of the LHC.

    We have generated a parameter space within our model satisfying various constraints coming from various theoretical considerations, electroweak precision observables, FCNS, 125 GeV Higgs signals, and searches of heavy resonances at the LHC.
    In our set-up, the right-handed doubly charged Higgs (pair produced) decays to a four-body final state via off-shell $W_R$ bosons, resulting in multi-lepton final states through the production of right-handed Majorana neutrinos. The multi-lepton signatures include four-lepton, six-lepton, and eight-lepton final states. However, the eight-lepton final state has very low statistics due to suppressed branching fractions of the doubly charged Higgs. Since the $\hpp$ decays to the four-body final state via the off-shell exchange of a heavy $W_R$ (5.56 TeV), its total decay width is quite small (${\cal O}(10^{-15})$ GeV) for a low mass ($150$ GeV). However, our analysis indicates that the $\hpp$ can decay within the detector of the LHC with a decay length of a few millimeters even for a heavier $W_R$ (10 TeV) as discussed in appendix~\ref{sec:decay-length-fWR}.

    We have studied the multi-lepton signatures of four-lepton, same-sign four-lepton, and six-lepton final states, over a mass range of 150 to 500 GeV for the $\hpp$ at the high luminosity phase of the LHC. Using a fast detector simulation of the monte-carlo generated events from various sources of SM background, we find that no background events survive in the same-sign four-lepton channel. In the six-lepton channel, we find that only a few background events survive the event selection criteria with an integrated luminosity of ${\cal L}=3000$ fb$^{-1}$. The six-lepton signatures for the signal can be observed with a good signal significance for $\hpp$ mass in the range of 200 to 500 GeV.  On the other hand, the SS4L signature gives a more striking and clear discovery channel for most of the right-handed doubly charged Higgs {mass range that we consider} in our leptophobic setup. 
    
   In addition, we also investigated the potential of future lepton colliders ($e^+e^-$ and muon collider) in detecting the multi-lepton collider signature of the $\hpp$. We find that both the future lepton colliders can probe the $\hpp$ in these multi-lepton signals better, compared to the LHC with a cleaner environment.  
   The multi-lepton signatures that we have investigated not only signify the presence of a triplet scalar but also imply the existence of a heavy charged gauge boson and heavy right-handed Majorana neutrinos.

\section*{Acknowledgment}	
The authors would like to acknowledge support from the Department of Atomic Energy, Government of India, for the Regional Centre for Accelerator-based Particle Physics (RECAPP), Harish-Chandra Research Institute.
	\appendix
	\section{Scalar potential and their mass matrices}\label{sec:potential}
	
	The most general  renormalizable Higgs potential invariant under the discrete parity and charge conjugation symmetries is given by \cite{Deshpande:1990ip,Dey:2015bur}
	\begin{eqnarray}
		\label{eq:scalar-potential}
		&&	V(\Phi,\Delta_{L,R},H_{L,R}) = \nonumber\\
		&& - \mu_1^2 {\rm Tr} (\Phi^{\dag} \Phi) - \mu_2^2
		\left[ {\rm Tr} (\tilde{\Phi} \Phi^{\dag}) + {\rm Tr} (\tilde{\Phi}^{\dag} \Phi) \right]
		- \mu_3^2 \left[ {\rm Tr} (\Delta_L \Delta_L^{\dag}) + {\rm Tr} (\Delta_R
		\Delta_R^{\dag}) \right] \nonumber
		\\
		&&-\mu_4^2  \left[H_R^\dagger H_R + H_L^\dagger H_L \right]
		+ \lambda_1 \left[ {\rm Tr} (\Phi^{\dag} \Phi) \right]^2 + \lambda_2 \left\{ \left[
		{\rm Tr} (\tilde{\Phi} \Phi^{\dag}) \right]^2 + \left[ {\rm Tr}
		(\tilde{\Phi}^{\dag} \Phi) \right]^2 \right\} \nonumber \\
		&&+ \lambda_3 {\rm Tr} (\tilde{\Phi} \Phi^{\dag}) {\rm Tr} (\tilde{\Phi}^{\dag} \Phi) +
		\lambda_4 {\rm Tr} (\Phi^{\dag} \Phi) \left[ {\rm Tr} (\tilde{\Phi} \Phi^{\dag}) + {\rm
			Tr}
		(\tilde{\Phi}^{\dag} \Phi) \right]\nonumber \\
		&& + \rho_1 \left\{ \left[ {\rm Tr} (\Delta_L \Delta_L^{\dag}) \right]^2 + \left[ {\rm
			Tr} (\Delta_R \Delta_R^{\dag}) \right]^2 \right\} 
		+ \rho_2 \left[ {\rm
			Tr} (\Delta_L \Delta_L) {\rm Tr} (\Delta_L^{\dag} \Delta_L^{\dag}) + {\rm Tr} (\Delta_R
		\Delta_R) {\rm Tr} (\Delta_R^{\dag} \Delta_R^{\dag}) \right] \nonumber
		\\
		&&+ \rho_3 {\rm Tr} (\Delta_L \Delta_L^{\dag}) {\rm Tr} (\Delta_R \Delta_R^{\dag})+
		\rho_4 \left[ {\rm Tr} (\Delta_L \Delta_L) {\rm Tr} (\Delta_R^{\dag} \Delta_R^{\dag}) +
		{\rm Tr} (\Delta_L^{\dag} \Delta_L^{\dag}) {\rm Tr} (\Delta_R
		\Delta_R) \right]  \nonumber \\
		&&+ \alpha_1 {\rm Tr} (\Phi^{\dag} \Phi) \left[ {\rm Tr} (\Delta_L \Delta_L^{\dag}) +
		{\rm Tr} (\Delta_R \Delta_R^{\dag})  \right] \nonumber
		\\
		&&+ \left\{ \alpha_2 e^{i \delta_2} \left[ {\rm Tr} (\tilde{\Phi} \Phi^{\dag}) {\rm Tr}
		(\Delta_L \Delta_L^{\dag}) + {\rm Tr} (\tilde{\Phi}^{\dag} \Phi) {\rm Tr} (\Delta_R
		\Delta_R^{\dag}) \right] + {\rm h.c.}\right\} \nonumber
		\\
		&&+ \alpha_3 \left[ {\rm Tr}(\Phi \Phi^{\dag} \Delta_L \Delta_L^{\dag}) + {\rm
			Tr}(\Phi^{\dag} \Phi \Delta_R \Delta_R^{\dag}) \right] + \beta_1 \left[ {\rm Tr}(\Phi
		\Delta_R \Phi^{\dag} \Delta_L^{\dag}) +
		{\rm Tr}(\Phi^{\dag} \Delta_L \Phi \Delta_R^{\dag}) \right] \nonumber \\
		&&+ \beta_2 \left[ {\rm Tr}(\tilde{\Phi} \Delta_R \Phi^{\dag} \Delta_L^{\dag}) + {\rm
			Tr}(\tilde{\Phi}^{\dag} \Delta_L \Phi \Delta_R^{\dag}) \right] + \beta_3 \left[ {\rm
			Tr}(\Phi \Delta_R \tilde{\Phi}^{\dag} \Delta_L^{\dag}) + {\rm Tr}(\Phi^{\dag} \Delta_L
		\tilde{\Phi} \Delta_R^{\dag}) \right] \nonumber\\
		&& +\lambda_H \left[ \left(H_R^\dagger H_R\right)^2  +\left(H_L^\dagger H_L\right)^2 \right] +\lambda_{H_{RL}} \left[ H_R^\dagger H_R H_L^\dagger H_L\right]\nonumber\\
		&&+\beta_H \left[ H_L^\dagger H_L {\rm Tr} (\Phi^{\dag} \Phi) + H_R^\dagger H_R{\rm Tr} (\Phi^{\dag} \Phi) \right]
		+ \eta_{H} \left[ H_R^\dagger H_R{\rm Tr} (\Delta_R^{\dag} \Delta_R) + H_L^\dagger H_L{\rm Tr} (\Delta_L^{\dag} \Delta_L) \right]\nonumber\\
		&&+ \eta_{H_{RL}} \left[ H_RH_R^\dagger{\rm Tr} (\Delta_L^{\dag} \Delta_L) + H_LH_L^\dagger{\rm Tr} (\Delta_R^{\dag} \Delta_R) \right]\nonumber\\
		&&+\eta_{H}^\prime \left[  {\rm Tr}(H_R^\dagger\Delta_R^\dagger\Delta_RH_R)-{\rm Tr}(H_R^\dagger\Delta_R\Delta_R^\dagger H_R)  +{\rm Tr}(H_L^\dagger\Delta_L^\dagger\Delta_LH_L)-{\rm Tr}(H_L^\dagger\Delta_L\Delta_L^\dagger H_L) \right]\nonumber\\
		&&+ \alpha_H \left[{\rm Tr}(H_L^\dagger\Phi\Delta_RH_R^\dagger)+{\rm Tr}(H_R^\dagger\Phi\Delta_LH_L^\dagger)   \right]\nonumber\\
		&&+\xi_H \left[{\rm Tr}(H_R\Delta_R^\dagger H_R) + {\rm Tr}(H_L\Delta_L^\dagger H_L)  + {\rm h.c.} \right]
		+ \xi_{H_{RL}}\left[ {\rm Tr}(H_R\Phi H_L^\dagger) + {\rm Tr}(H_L\Phi H_R^\dagger) + {\rm h.c.}  \right].
	\end{eqnarray}
	The neutral scalar fields in the above potential can be expressed in terms of their $CP$-even and -odd components:
	\begin{subequations}\label{eq:neutral-scalar}
		\begin{align}
			\phi_1^0 &= \frac{1}{\sqrt{2}} \left(v_1 + \sigma_1 + i \varphi_1 \right)\,, &  \delta_L^0 &= \frac{1}{\sqrt{2}} \left(v_{tL} + \sigma_L + i \varphi_L \right)\,,\\
			\phi_2^0 &= \frac{1}{\sqrt{2}} \left(v_2 + \sigma_2 + i \varphi_2 \right)\,,  & \delta_R^0 &= \frac{1}{\sqrt{2}} \left(v_{tR} + \sigma_R + i \varphi_R \right)\,, \\
			H_L^0 &= \frac{1}{\sqrt{2}} \left(v_{L} + \sigma_{H_L} + i \varphi_{H_L} \right)\,,  & H_R^0 &= \frac{1}{\sqrt{2}} \left(v_{R} + \sigma_{H_R} + i \varphi_{H_R} \right)\,,
		\end{align}
	\end{subequations}
	where we use the generic symbols $\sigma$ and $\varphi$ to label the $CP$-even and -odd states, respectively. 
	For the vacuum expectation values, which we assume to be real, we use the following parametrization:
	\begin{align}
		v_1 &= v \cos \beta\,,\qquad v_2 = v \sin\beta\,, \qquad t_\beta \equiv \tan\beta=\frac{v_2}{v_1}\,,
		\qquad v = v_1^2 + v_2^2\,.
	\end{align}
 The EW vev is then  given by $v_{\rm EW}=\sqrt{ v^2+v_{tL}^2+v_L^2 }$.
	For the exact LR symmetry, we consider the $SU(2)$ gauge coupling to be the same, i.e., $g_R=g_L$. The six neural scalar acquiring vevs  provide six minimization conditions, 
 \begin{equation}\label{eq:tadpole}
 \frac{\partial }{\partial S_i^0}V(\Phi,\Delta_{L,R},H_{L,R})|_{\langle S_i^0 \rangle}=0,~~~S_i=\{\phi_1,\phi_2,\delta_{L,R},h_{L,R}\}.
 \end{equation}
 Solving the six minimization conditions (tadpole) for the potential, we eliminate the following six parameters: $\mu_1$, ~$\mu_2$,~ $\mu_3$,~ $\mu_4$,~ $\beta_2$ and $\lambda_{H_{RL}}$. In the limit of $v_2=0$, $v_L\simeq 0$, $v_{tL}\simeq 0$, and $\xi_{H_{RL}}=0$, the solutions for tadpole equations are given by,
 \begin{eqnarray}\label{eq:tadpol}
\mu_1^2&=& \frac{1}{2} \left(-\alpha_1 v_{tR}^2-\beta_H v_{R}^2-2 \lambda_1
v_1^2\right),\nonumber\\\
\mu_2^2&=& \frac{1}{2} \left(-\alpha_2 v_{tR}^2-\lambda_4
v_1^2\right),\nonumber\\
\mu_3^2&=& \frac{1}{2} \left(-\alpha_1 v_1^2+v_{R}^2
(\eta_H^\prime-\eta_H)-2 \rho_1 v_{tR}^2+\frac{\sqrt{2} v_{R}^2
	\xi_H}{v_{tR}}\right),\nonumber\\
\mu_4^2&=& \frac{1}{2} \left(-\beta_H v_1^2+v_{tR} \left(\eta_H^\prime
v_{tR}-\eta_H v_{tR}+2 \sqrt{2} \xi_H\right)-2 \lambda_H v_{R}^2\right),\nonumber\\
\beta_2&=& 0,\nonumber\\
\lambda_{H_{RL}}&=&
\lambda_H-\frac{v_{tR} \left(v_{tR} (\eta_H^\prime+\eta_{H_{RL}}-\eta_H)+2 \sqrt{2} \xi_H\right)}{2
	v_{R}^2}.
\end{eqnarray}	

 In the limit of $v_2=0$, $v_L\simeq 0$, $v_{tL}\simeq 0$, and $\xi_{H_{RL}}=0$, the squared mass matrix for the neutral $CP$-even scalars, $CP$-odd  pseudo scalars, and singly charged scalars arranged in the basis $\left(\phi_1,\phi_2,h_L,\delta_L,h_R,\delta_R \right)$ are given by,
\begin{eqnarray}
m_H^2&=&\left\{\left\{2 \lambda_1 v_1^2,-2 \lambda_4 v_1^2,0,0,\beta_H v_1 v_{HR},\alpha_1
v_1 v_{tR}\right\},\right.\nonumber\\
&&\left\{-2 \lambda_4 v_1^2,2 (2 \lambda_2+\lambda_3) v_1^2+\frac{\alpha_3 v_{tR}^2}{2},-\frac{1}{2} \alpha_H v_{HR}
v_{tR},\frac{\beta_1 v_1 v_{tR}}{2},0,-2 \alpha_2 v_1 v_{tR}\right\},\nonumber\\
&&\left\{0,-\frac{1}{2} \alpha_H v_{HR}
v_{tR},0,-\frac{1}{2} \alpha_H v_1 v_{HR},0,0\right\},\nonumber\\
&&\left\{0,\frac{\beta_1 v_1 v_{tR}}{2},-\frac{1}{2} \alpha_H v_1 v_{HR},\frac{(\eta_H^\prime+\eta_{H_{RL}}-\eta_H)
v_{HR}^2 v_{tR}+(-2 \rho_1+\rho_3) v_{tR}^3+\sqrt{2} v_{HR}^2 \xi_H}{2 v_{tR}},0,0\right\},\nonumber\\
&&\left\{\beta_H v_1 v_{HR},0,0,0,2 \lambda_H v_{HR}^2,-v_{HR} \left(\eta_H^\prime v_{tR}-\eta_H v_{tR}+\sqrt{2}
\xi_H\right)\right\},\nonumber\\
&&\left.\left\{\alpha_1 v_1 v_{tR},-2 \alpha_2 v_1 v_{tR},0,0,-v_{HR} \left(\eta_H^\prime v_{tR}-\eta_H
v_{tR}+\sqrt{2} \xi_H\right),2 \rho_1 v_{tR}^2+\frac{v_{HR}^2 \xi_H}{\sqrt{2} v_{tR}}\right\}\right\}
,\end{eqnarray}
\begin{eqnarray}
m_A^2&=&\left\{\bigg\{0,0,0,0,0,0\bigg\},\left\{0,2 (-2 \lambda_2+\lambda_3) v_1^2+\frac{\alpha_3 v_{tR}^2}{2},-\frac{1}{2}
\alpha_H v_{HR} v_{tR},\frac{\beta_1 v_1 v_{tR}}{2},0,0\right\}\right.,\nonumber\\
&&\left\{0,-\frac{1}{2} \alpha_H v_{HR} v_{tR},0,-\frac{1}{2}
\alpha_H v_1 v_{HR},0,0\right\},\nonumber\\
&&\left\{0,\frac{\beta_1 v_1 v_{tR}}{2},-\frac{1}{2} \alpha_H v_1 v_{HR},\frac{(\eta_H^\prime+\eta_{H_{RL}}-\eta_H)
v_{HR}^2 v_{tR}+(-2 \rho_1+\rho_3) v_{tR}^3+\sqrt{2} v_{HR}^2 \xi_H}{2 v_{tR}},0,0\right\},\nonumber\\
&&\left.\left\{0,0,0,0,2 \sqrt{2} v_{tR} \xi_H,-\sqrt{2} v_{HR} \xi_H\right\},\left\{0,0,0,0,-\sqrt{2} v_{HR} \xi_H,\frac{v_{HR}^2
\xi_H}{\sqrt{2} v_{tR}}\right\}\right\},
\end{eqnarray}
and
\begin{eqnarray}
m_{H^{\pm}}^2&=&\left\{\bigg\{0,0,0,0,0,0\bigg\},
\left\{0,\frac{\alpha_3 v_{tR}^2}{2},-\frac{1}{2} \alpha_H v_{HR} v_{tR},0,\frac{\beta_1
v_1 v_{tR}}{2 \sqrt{2}},\frac{\alpha_3 v_1 v_{tR}}{2 \sqrt{2}}\right\}\right.,\nonumber\\
&&\left\{0,-\frac{1}{2} \alpha_H v_{HR} v_{tR},0,0,-\frac{\alpha_H
v_1 v_{HR}}{2 \sqrt{2}},-\frac{\alpha_H v_1 v_{HR}}{2 \sqrt{2}}\right\},\nonumber\\
&&\left\{0,0,0,\frac{1}{2} (\eta_H^\prime+\eta_H^\prime) v_{tR}^2+\sqrt{2} v_{tR} \xi_H,0,-\frac{1}{4} v_{HR} \left(\sqrt{2} (\eta_H^\prime+\eta_H^\prime)
v_{tR}+4 \xi_H\right)\right\},\nonumber\\
&&\left\{0,\frac{\beta_1 v_1 v_{tR}}{2 \sqrt{2}},-\frac{\alpha_H v_1 v_{HR}}{2 \sqrt{2}},0,\frac{1}{4 v_{tR}}\bigg(\alpha_3 v_1^2
v_{tR}+2 \left((\eta_H^\prime+\eta_{H_{RL}}-\eta_H) v_{HR}^2 v_{tR}\right.\right.\nonumber\\
&&\left.\left.-(2 \rho_1-\rho_3) v_{tR}^3+\sqrt{2} v_{HR}^2\xi_H\right)\bigg),\frac{\beta_1 v_1^2}{4}\right\},\nonumber\\
&&\left.\left\{0,\frac{\alpha_3 v_1 v_{tR}}{2 \sqrt{2}},-\frac{\alpha_H v_1 v_{HR}}{2 \sqrt{2}},-\frac{1}{4} v_{HR}
\left(\sqrt{2} (\eta_H^\prime+\eta_H^\prime) v_{tR}+4 \xi_H\right),\frac{\beta_1 v_1^2}{4},\right.\right.\nonumber\\
&&\left.\left.\frac{1}{{4 v_{tR}}}\bigg(\alpha_3 v_1^2 v_{tR}+v_{HR}^2
\left((\eta_H^\prime+\eta_H^\prime) v_{tR}+2 \sqrt{2} \xi_H\right)\bigg)\right\}\right\}.
\end{eqnarray}

 \section{Decay width and decay length of $\hpp$ for heavier $W_R$}\label{sec:decay-length-fWR}
	\begin{figure*}
		\centering
		\includegraphics[width=0.49\textwidth]{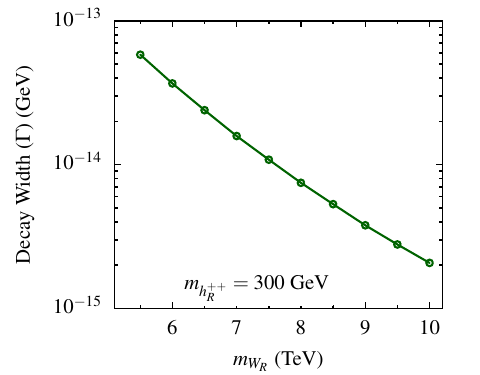}
		\includegraphics[width=0.49\textwidth]{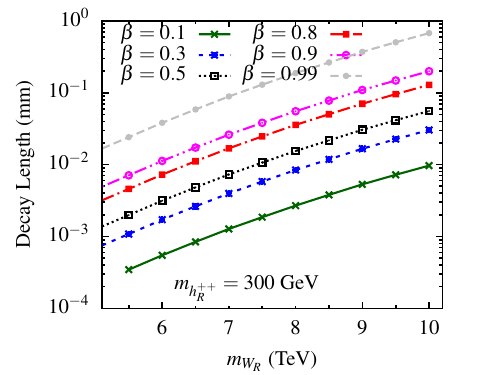}
		\caption{\label{fig:decayhRpp-vs-WR} Total decay width ({\em left-panel}) and decay length ({\em right-panel}) for various choice of boost $\beta$ of $h_R^{++}$ as a function of $W_R$ mass for  $\mhpp=300$ GeV.}
	\end{figure*}
	The $\hpp$  in our model has its primary decay through off-shell $W_R$ as its couplings to leptons are tiny. This would lead to a very small decay width and hence a large lifetime if the off-shell $W_R$ mediating the four-body decay (see Fig.~\ref{fig:Feynman-diags}) become very heavy. Here, we calculate the decay width and decay length of $\hpp$ for $W_R$ mass heavier than $5.5$ TeV, which is considered as a benchmark choice, for $\mhpp=300$ GeV, as an example. The total decay width and decay length for a few choices of boost $\beta$ of $\hpp$ are shown in Fig.~\ref{fig:decayhRpp-vs-WR} as a function of $W_R$ mass in the range $\sim 5.5$ - $10$ TeV. The decay width decreases and decay length increases with an increase in the $W_R$ mass. However, the decay length remains within a few mm even	for $m_{W_R}\approx 10$ TeV and $\beta\to 1$ for $\mhpp=300$ GeV. We have checked that the decay length has a similar value for all our choices of $\mhpp$ in this work.
\section{Doubly charged scalar's contribution to $h\to \gamma\gamma$}\label{sec:higgs-to-diphoton}
The charged Higgs (singly and doubly) modify the Higgs to di-photon partial decay width via a triangular loop as they couple to both the photon and the SM Higgs boson. The modification factor, i.e., the ratio between the partial decay width of the Higgs to di-photon in a new physics model to that in the SM 
can be expressed as~\cite{Martinez:1989bg,Picek:2012ei,Carena:2012xa,Bambhaniya:2013wza},
\begin{equation}
 R_{\gamma\gamma} =  \left|1 + \sum_{S=h^{\pm \pm},h^{\pm}} Q_S^2 \frac{c_S}{2} \frac{v^2}{m_S^2} 
 \frac{A_0(\tau_S)}{A_1(\tau_{W_L}) + N_c Q_t^2 A_{1/2}(\tau_t)} \right|^2.
\end{equation}
Here, $Q_S$ is the electric charge  in units of $e$,  $m_S$ is a mass,  $N_c$ is the color factor of charged scalars; and $\tau_{i} = 4 m_i^2/m_{h}^2 (i = W_{L},t,S)$.
The quantity $c_S$ is the coupling of the Higgs boson with the charged scalars.
The couplings $c_S$s are given by
\begin{eqnarray}
c_{hh^+h^-} =-\Biggl[\frac{2 {\alpha_1} v^2+8 {\alpha_2} {v_1} {v_2}+{\alpha_3} (v^2)}{2 {v^2}}\Biggr] ,\label{cs1}\nonumber\\
c_{hh^{++}h^{--}} =-\Biggl[\frac{{\alpha_1} v^2+{v_1} (4 {\alpha_2} {v_2}+{\alpha_3} {v_1})}{{v^2} }\Biggr]. \label{cs2}
\end{eqnarray}
In our benchmark scenario, $h^+$ and $h_L^{++}$ are heavy and thus do not contribute to the Higgs to di-photon decay.  Additionally, we choose $\alpha_{1,2}=0$ and $v_2= 0$ leaving only the $\alpha_3$ contributing to the Higgs to di-photon decay branching ratio.
The loop functions $A_{1/2}$, $A_{1}$ and $A_0$ corresponding to fermions, vector bosons, and scalars respectively, are given by
\begin{eqnarray}
 A_{1/2}(x) &=& 2x^2[x^{-1}+(x^{-1}-1)f(x^{-1})], \nonumber\\
A_{1}(x) &=& -x^2[2x^{-2}+3x^{-1}+3(2x^{-1}-1)f(x^{-1})], \nonumber\\
A_{0}(x) &=& -x^2[x^{-1}-f(x^{-1})],
\end{eqnarray}
with $f(x) = \left(\sin^{-1}(\sqrt{x})\right)^2$ for $m_h<2m_{\rm loop}$.

	\bibliographystyle{utphysM}
	\bibliography{References}
\end{document}